\documentclass[lettersize,journal]{IEEEtran}

\usepackage{amsmath}
\usepackage{tabularx}
\usepackage{fontawesome}
\usepackage{xargs} 
\usepackage{xspace}
\usepackage{multirow}
\usepackage{enumerate}
\usepackage{enumitem}
\usepackage{makecell}
\usepackage{soul}

\usepackage{booktabs}
\usepackage{xcolor}
\usepackage{colortbl}
\usepackage{array}

\definecolor{lightgray}{gray}{0.95}

\usepackage{graphicx}
\usepackage{subcaption}   
\usepackage{hyperref}     
\usepackage[nameinlink,capitalise,noabbrev]{cleveref}

\usepackage{pifont} 
\newcommand{\cmark}{\textcolor{green!60!black}{\ding{51}}} 
\newcommand{\xmark}{\textcolor{red!70!black}{\ding{55}}}   

\definecolor{highred}{HTML}{FF403F} 
\definecolor{midgold}{HTML}{FFE13F} 
\definecolor{lowblue}{HTML}{56ACFF} 

\usepackage[nolist,printonlyused]{acronym}
\acrodef{PDF}{Probability Density Function}
\acrodef{EPS}{Evolved Packet System}
\acrodef{TTI}{Transmission Time Interval}
\acrodef{UTLA}{Upper Tier Local Authority}
\acrodef{LAD}{Local Authority District}
\acrodef{CA}{Carrier Aggregation}
\acrodef{BSC}{Base Station Controller}
\acrodef{EPC}{Evolved Packet Core}
\acrodef{PDP}{Packet Data Protocol}
\acrodef{GMM}{GPRS Mobility Management}
\acrodef{GGSN}{Gateway GPRS Support Node}
\acrodef{NAS}{Non-Access Stratum}
\acrodef{RAN}{Radio Access Network}
\acrodef{GPRS}{General Packet Radio System}
\acrodef{UTRAN}{UMTS Terrestrial Radio Access Network}
\acrodef{SGSN}{Serving GPRS Support Node}
\acrodef{CN}{Core Network}
\acrodef{RNC}{Radio Network Controller}
\acrodef{MME}{Mobility Management Entity}
\acrodef{UMTS}{Universal Mobile Telecommunication Systems}
\acrodef{xDR}{eXtended Detail Record}
\acrodef{TAC}{Type Allocation Code}
\acrodef{PoP}{Point of Presence}
\acrodef{IMEI}{International Mobile Equipment Identity}
\acrodef{EMSISDN}{Encrypted Mobile Station International Subscriber Directory Number}
\acrodef{MSISDN}{Mobile Station International Subscriber Directory Number}
\acrodef{VoLTE}{Voice over \ac{LTE}}
\acrodef{MVNO}{Mobile Virtual Network Operators}
\acrodef{CDR}{Call Detail Record}
\acrodef{IMSI}{International Mobile Subscriber Identity}
\acrodef{RAT}{Radio Access Technology}
\acrodef{LPWA}{Low Power Wide Area}
\acrodef{VMNO}{Visited Mobile Network Operator}
\acrodef{HMNO}{Home Mobile Network Operator}
\acrodef{DCH}{Data Clearing House}
\acrodef{GSMA}{\ac{GSM} Association}
\acrodef{GSM}{Global System for Mobile communications}
\acrodef{TAP}{Transferred Account Procedure}
\acrodef{LTE-M}{\ac{LTE} Machine Type Communication}
\acrodef{MTC}{Machine Type Communications}
\acrodef{NB-IoT}{Narrow Band \ac{IoT}}
\acrodef{IOT}{Inter Operator Tariff}
\acrodef{IoT}{Internet of Things}
\acrodef{M2M}{Machine-to-Machine}
\acrodef{FNO}{Fixed Network Operator}
\acrodef{ISP}{Internet Service Providers} 
\acrodef{IPX}{IP Packet Exchange}
\acrodef{2G}{Second Generation}
\acrodef{3G}{Third Generation}
\acrodef{4G}{Fourth Generation}
\acrodef{5G}{Fifth Generation}
\acrodef{ADB}{Android Debug Bridge}
\acrodef{ASCI}{Advertising Standards Council of India}
\acrodef{ASN}{Autonomous System Number}
\acrodef{CDN}{Content Delivery Network}
\acrodef{DL}{Downlink}
\acrodef{DNS}{Domain Name Service}
\acrodef{EaaS}{Experiment as a Service}
\acrodef{ECDF}{Empirical Cumulative Distribution Function}
\acrodef{HTTP}{Hyper Text Transfer Protocol}
\acrodef{ICMP}{Internet Control Message Protocol}
\acrodef{ISP}{Internet Service Provider}
\acrodef{IXP}{Internet Exchange Point}
\acrodef{LTE}{Long Term Evolution}
\acrodef{E2E}{End-to-end}
\acrodef{QCI}{QoS Class Identifier}
\acrodef{NDT}{Network Diagnostic Tool}
\acrodef{QoE}{Quality of Experience}
\acrodef{QoS}{Quality of Service}
\acrodef{OS}{Operating System}
\acrodef{APN}{Access Point Name}
\acrodef{RMBT}{RTR Multithreaded Broadband Test}
\acrodef{RTT}{Round-Trip Time}
\acrodef{SIM}{Subscriber Identity Module}
\acrodef{SIMs}{Subscriber Identity Modules}
\acrodef{SLA}{Service-Level Agreement}
\acrodef{TCP}{Transmission Control Protocol}
\acrodef{UDP}{User Datagram Protocol}
\acrodef{UE}{User Equipment}
\acrodef{UL}{Uplink}
\acrodef{NRA}{National Regulatory Authority}
\acrodef{EC}{European Commission}
\acrodef{SGW}{Serving Gateway}
\acrodef{PGW}{Packet Data Network Gateway}
\acrodef{GTP}{GPRS Tunneling Protocol}
\acrodef{MNO}{Mobile Network Operator}
\acrodef{non-IID}{non-independent and identically distributed}
\acrodef{EU}{European Union}
\acrodef{HR}{home-routed roaming}
\acrodef{LBO}{local breakout}
\acrodef{IHBO}{IPX hub breakout}
\acrodef{VoIP}{Voice over IP}
\acrodef{IPG}{Inter-Packet Gap}
\acrodef{KS}{Kolmogorov-Smirnov}
\acrodef{FQDN}{Fully Qualified Domain Name}
\acrodef{MNC}{Mobile Network Code}
\acrodef{MCC}{Mobile Country Code}
\acrodef{MCCMNC}{\ac{MCC}-\ac{MNC}}
\acrodef{SIP}{Session Initiation Protocol}
\acrodef{g}{radius of gyration, (or gyration for short)}
\acrodef{MBB}{Minimum Bounding Box}
\acrodef{BBD}{Minimum Bounding Box Diagonal}
\acrodef{KPI}{Key Performance Indicator}
\acrodef{e}{Entropy}
\acrodef{FL}{Federated Learning}
\acrodef{RoG}{Radius of Gyration}
\acrodef{ML}{Machine Learning}
\acrodef{IID}{independent and identically distributed}
\acrodef{BS}{Base Station}
\acrodef{NSA}{Non-Stand Alone}
\acrodef{SA}{Stand Alone}
\acrodef{POI}{Point of Interest}
\acrodefplural{POI}[POIs]{Points of Interest}
\acrodef{AMF}{Access and Mobility Management Function}
\acrodef{RNN}{Recurrent Neural Network}
\acrodef{MSE}{Mean Squared Error}
\acrodef{FedAvg}{Federated Averaging}
\acrodef{FedProx}{Federated Proximal}
\acrodef{5GC}{5G Core}
\acrodef{MSC}{Mobile Switching Center}

\newcommand{\eg}{\textit{e.g.},\xspace}
\newcommand{\ie}{\textit{i.e.},\xspace}

\renewcommand\sectionautorefname{\S\kern-2pt}
\renewcommand\subsectionautorefname{\S\kern-2pt}
\renewcommand\subsubsectionautorefname{\S\kern-2pt}

\begin{document}

\title{Device-Induced Mobility Heterogeneity and Its Impact on Federated Next-Location Prediction}

\author{
André~Felipe~Zanella,
Paula~Delgado~de~Santos,
José~Suárez-Varela,
David~Solans~Noguero,
Andra~Lutu,
and~Jesus~Omaña-Iglesias


\thanks{All authors are with Telefonica Research.}

} 



\maketitle

\begin{abstract}
Predicting human mobility is central to network management, urban planning, and location-based services, yet privacy concerns are increasingly pushing such tasks toward federated learning. A key challenge in this setting is the strong heterogeneity of mobility data across users, which leads to non-IID data distributions and can significantly impact model performance and fairness.
In this paper, we present a large-scale measurement study of device-level mobility heterogeneity using data from a major \ac{MNO} in a dense European capital. Our dataset comprises mobility traces derived from network events, covering over $50{,}000$ UEs across 178 smartphone models. We group devices into tiers based on their market segment and show that mobility characteristics, such as spatial range and regularity, vary systematically across these tiers, revealing a previously underexplored source of bias in mobility data.
We investigate the implications of this heterogeneity on the use case of federated next-location prediction. In particular, we find that balancing device tiers across silos through tier-aware client selection yields the best results, improving fairness by up to 34.49\% without degrading accuracy. In contrast, a FedProx-only approach achieves a 21.46\% fairness improvement, offering a reasonable alternative when device information is unavailable, though with lower overall gains.
\end{abstract}

\begin{IEEEkeywords}
Handovers, Mobility, Federated Learning, Next-location prediction.
\end{IEEEkeywords}

\section{Introduction}
\label{sec:intro}
Understanding human mobility patterns is fundamental to applications such as network management, urban planning, and location-based services. Large-scale mobility datasets, often collected from mobile devices through network events (CDR, handovers) or application logs (GPS), have enabled the development of predictive models by leveraging the temporal and spatial regularities of human movement~\cite{gonzalez2008understanding,song2010limits}.

However, while mobile devices serve as the primary source of mobility data, the role of device diversity in shaping observed mobility patterns remains poorly understood. Modern datasets are inherently device-mediated: differences in smartphone characteristics, usage, and adoption may influence both how mobility is captured and which parcel of population is represented. As a result, device heterogeneity may introduce systematic biases in mobility datasets, with direct implications for downstream analysis and modeling.

In this paper, we present a large-scale measurement study of device-level mobility heterogeneity using data from a major \ac{MNO} in a dense European capital. Our dataset consists of mobility traces derived from network handover events, covering over $50{,}000$ \acp{UE} across 178 smartphone models. We group devices into tiers based on their characteristics, and analyze how mobility characteristics vary across these groups. We note that mobility patterns differ systematically across device tiers, revealing a previously underexplored source of bias in mobility data.

Based on this knowledge, we further investigate the implications of this device-induced mobility heterogeneity for \ac{FL}, a paradigm increasingly adopted for privacy-preserving mobility modeling~\cite{belal2024survey}. While prior work has primarily treated device heterogeneity as a systems issue, \eg removing lower-capability \acp{UE} to mitigate the effect of stragglers~\cite{yang2018applied}, our findings highlight its role as a data bias problem: imbalances in device populations lead to skewed mobility distributions, which in turn affect model performance and fairness. Through controlled experiments on federated next-location prediction, we show that device-induced heterogeneity significantly impacts fairness across users and devices when not properly addressed.

Our contributions are as follows:
\begin{itemize}
    \item We conduct a large-scale empirical analysis of mobility traces from $50{,}000+$ \acp{UE} across 178 smartphone models, and categorize devices into capability tiers, revealing a strong skew toward high-tier devices~(\S\ref{sec:system_analysis}).

    \item We demonstrate that Entropy, \ac{RoG}, and weekly temporal patterns differ significantly across smartphone capability tiers, and that these tiers align closely with independently derived mobility clusters, revealing smartphone models as a systematic proxy of mobility behavior~(\S\ref{sec:mobility_analysis}).

    \item We demonstrate that device-induced data heterogeneity impacts fairness in federated next-location prediction, and show that tier-aware client selection improves fairness by up to 34.49\% without sacrificing accuracy; when device information is unavailable, FedProx can be an alternative for improving fairness, (+21.46\%), altough not as effective as balancing client selection across tiers~(\S\ref{sec:eval},\S\ref{sec:results}).

\end{itemize}

\section{Related Work}
\label{sec:relwork}

The growing interest in \ac{FL} for next-location prediction is largely driven by privacy concerns surrounding centralized mobility datasets~\cite{belal2024survey,gecer2024federated}. At the same time, human mobility is inherently heterogeneous, ranging from highly routine to exploratory behaviors, with exploratory users being significantly harder to predict~\cite{amichi2020understanding,amichi2022revealing}. This heterogeneity translates into strongly non-IID trajectory data, posing fundamental challenges for federated training.

Prior work has explored federated approaches for human and vehicular mobility prediction, including privacy-preserving forecasting with personalized local adaptors~\cite{feng2020pmf}, reinforcement learning-based trajectory prediction~\cite{li2025fedddpg}, and clustering-based methods that group clients according to mobility patterns~\cite{zhang2023federated}. Other approaches prioritize selecting high-entropy users with the goal of improving model generalization~\cite{park2023fedgeo}. While effective for improving system accuracy, these methods typically rely on additional preprocessing steps to extract mobility characteristics prior to training, and treat heterogeneity primarily as a modeling and accuracy challenge, without exploring fairness. Similarly, prior work has shown that high heterogeneity across FL silos can reduce model accuracy~\cite{esper2024beauty}, yet the underlying sources of this heterogeneity remain insufficiently understood.

A key limitation of existing work lies in the datasets used for mobility modeling. Common benchmarks such as Foursquare~\cite{foursquare_dataset}, Gowalla~\cite{gowalla_dataset}, and Yelp~\cite{yelp_dataset} provide rich spatial context (\eg points of interest), but lack information about the devices or system-level conditions under which data is generated. This absence of device-level context limits the ability to study how data collection mechanisms and user populations influence observed mobility patterns, and prevents the analysis of device-induced biases. Evidence from federated recommendation systems suggests that excluding low-tier devices can significantly degrade model fairness~\cite{maeng2022towards}. However, such exclusions are common in \ac{FL} to optimize system performance, as stragglers can increase training time by $2.32\times$ and reduce accuracy by up to $9.2\%$~\cite{yang2021characterizing}, leading to the exclusion of lower-capability \acp{UE} as a common mitigation strategy~\cite{yang2018applied}. This highlights the importance of understanding how device populations shape the data distribution, and how this can affect distributed learning tasks.

In parallel, recent measurement studies have examined device heterogeneity in mobile networks. Control-plane data from the \ac{MME} provides a network-wide view of \ac{UE} behavior, revealing the diversity of connected devices and their evolution across radio access technologies~\cite{kalntis2024through,parastar2023spotlight}. These works show that smartphones dominate network traffic, while \ac{M2M} and legacy devices still represent a non-negligible share. However, they provide only coarse-grained views of mobility and do not analyze how mobility patterns systematically vary across device types.

Taken together, these gaps point to a missing link between mobile network measurements and federated mobility modeling, an important matter as optimizing system performance may inadvertently exacerbate data bias. Understanding this coupling is essential for designing \ac{FL} systems that are both efficient and fair. In this work, we bridge this gap by performing a large-scale measurement campaign in \ac{MNO} mobility events,  quantifying how device heterogeneity translates into mobility heterogeneity, and by evaluating its impact on fairness and performance in federated next-location prediction.

\section{Data and Methods}
\label{sec:methods}
In this section, we detail our measurements and processing methodology, from collecting handover events, creating mobility trajectories and computation mobility indicators.

\begin{figure*}
\centering
    \begin{subfigure}[t]{0.32\textwidth}
        \centering
        \includegraphics[height=0.42\columnwidth]{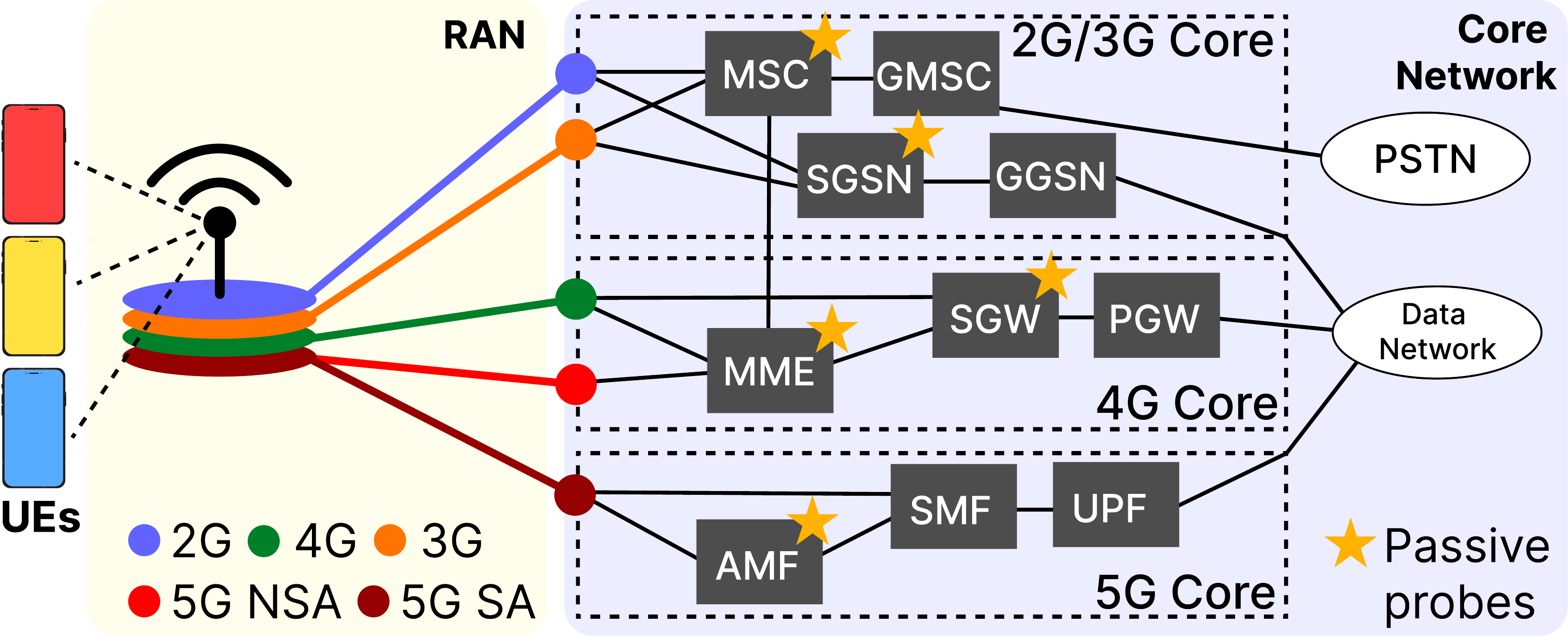}
        \caption{}
        \label{fig:network_probes}
    \end{subfigure}
    \hspace{0.1cm}
    \begin{subfigure}[t]{0.32\textwidth}
        \centering
        \includegraphics[height=0.42\columnwidth]{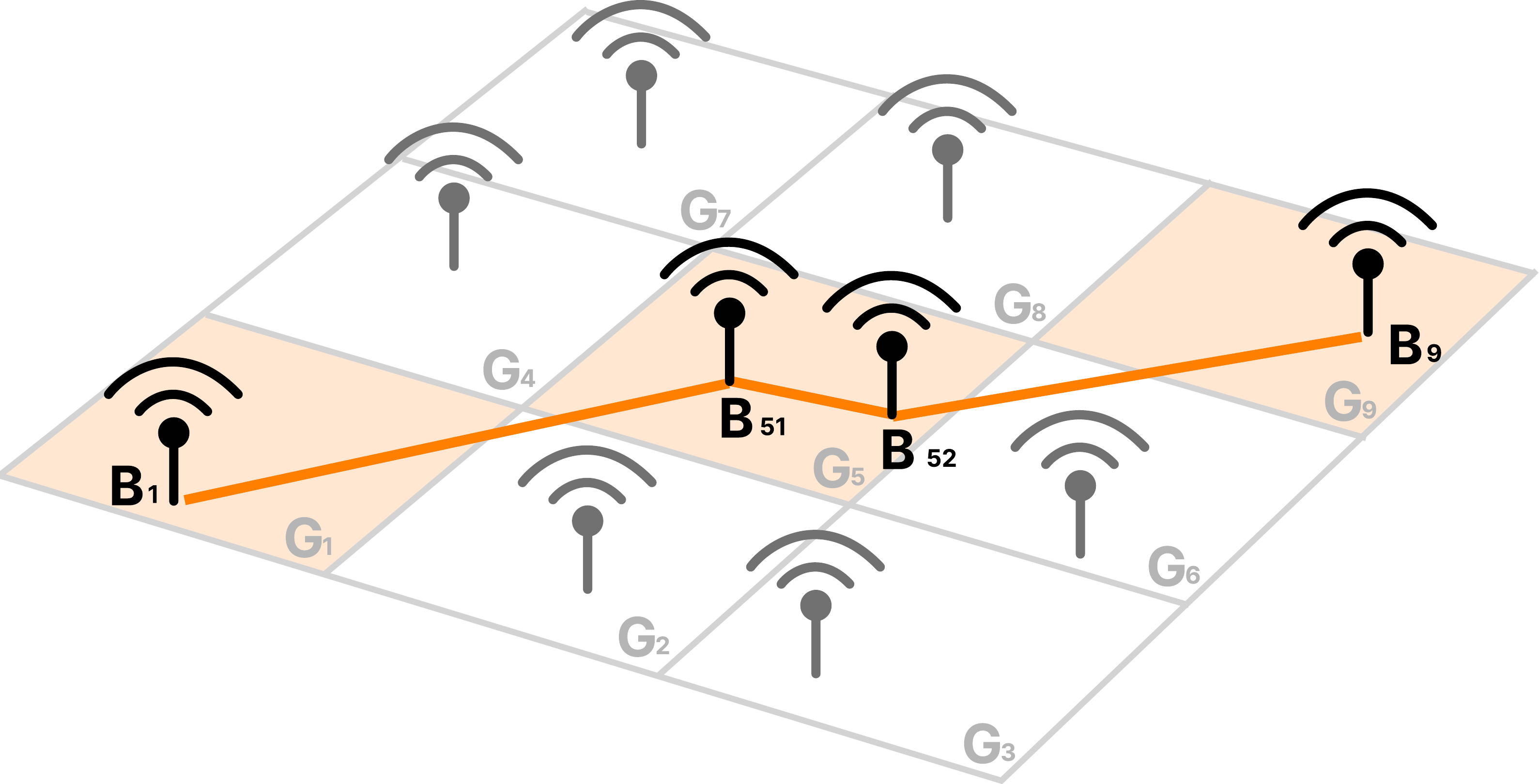}
        \caption{}
        \label{fig:network_grid_example}
    \end{subfigure}
    \begin{subfigure}[t]{0.32\textwidth}
        \centering
        \includegraphics[
            height=0.42\columnwidth,
            trim={1.4cm 3.28cm 1.35cm 3cm},
            clip
        ]{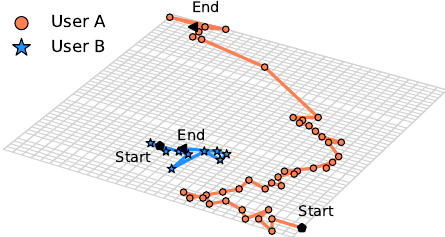}
        \caption{}
        \label{fig:example_trajectories}
    \end{subfigure}
    \caption{
    (a) High-level overview of measurements. 
    (b) Overview of the conversion from base stations to grid. 
    (c) Trajectories from two test devices with differences in length and diversity, reflected on mobility statistics. User A (orange) - Entropy: 4.76, RoG: 0.12; User B (blue) - Entropy: 1.68, RoG: 0.01.
    }
\label{fig:network_examples}
\end{figure*}

\subsection{Network measurements infrastructure}
\label{subsec:network_infra}

The mobile network we study supports from 2G to 5G \acp{RAT}. \Cref{fig:network_probes} highlights the vantage points we leverage from the architecture.
We simplify the network representation into three main domains: (i) the \ac{UE} (\eg smartphones), (ii) the \ac{RAN} and (iii) the \ac{CN}.
Our passive measurement approach relies on commercial solutions the \ac{MNO} integrates within its infrastructure. 
In particular, we rely on passive measurements collected from the \ac{MSC} and the \ac{SGSN} (2G \& 3G connectivity), the \ac{MME} in the \ac{EPC} (4G \& 5G NSA) and the \ac{AMF} in the \ac{5GC} (5G SA), respectively. We collect control plane information related to mobility events from the population of devices connected to the \ac{MNO}’s \ac{RAN}.

\subsection{Data feeds}
\label{subsec:data-feeds}
We rely on various data feeds from the measurement infrastructure, collected at \ac{BS} level and later aggregated at grid level. 

\textit{General Signaling Dataset.}
We capture the activity of control plane devices across \acp{RAT} supported by the cellular provider. For every \ac{RAT}, the collected dataset includes the (control plane) signaling messages related to events triggered by the \ac{MNO}'s subscribers (\eg Attach, Handover and Detach).
Each event carries anonymized device-level identifiers, including an anonymized \ac{IMEI}, the radio sector ID handling the communication, timestamp, and event result code (success / failure).

\textit{Devices Catalog.}
We use a commercial database provided by \ac{GSMA} which maps the device \ac{TAC} (the first 8 digits of the \ac{IMEI}) to a set of properties (\eg brand, model, radio connectivity). This allows us to distinguish between smartphones (likely used as primary devices by the mobile users) and \ac{M2M} devices. We also use this information to associate multiple \ac{TAC} values that belong to the same smartphone model (\eg due to regional variations). 

\textit{Radio Network Topology.}
To account for potential structural changes in the radio access network (\eg new site deployments), we rely on a daily snapshot of the network topology. This includes metadata (location  and configuration) and the status (active/inactive) of each \ac{BS}.  

\textit{Ethics statement.}
The datasets used in this research are protected by Non-Disclosure Agreements (NDAs) that prohibit sharing with unauthorized parties or public repositories. Data collection and storage procedures strictly adhere to the MNO’s guidelines and comply with local regulations. Although some metrics are computed at the user level, all data handling is limited to producing aggregated, anonymized insights, with no access to individual locations or trajectories. No personal or contractual information was available, and none of the authors were involved in raw data extraction or encryption. As such, the study poses no risk to mobile subscribers while offering valuable insights into mobility patterns.

\subsection{Trajectories processing}
\label{subsec:trajectories_processing}

We take as reference a 14-day period between Jul. 22 and Aug. 4, 2024, and rely on anonymized trajectories across unique \acp{UE} (containing only native \ac{MNO} subscribers and no international roamers). Specifically, the sample includes data from all the \acp{UE} signaling over \ac{BS} within the studied metropolitan area for at least 80\% of the time during the 14-day period. 
We filter out all \acp{UE} that have their \ac{TAC} value not associated with a smartphone model, and keep only \acp{UE} with $\geq$10 active days and $\geq$120 total records~\cite{esper2024beauty}, excluding smartphone models with $<$20 unique \acp{UE}. 

Using the signaling data, we represent each anonymized \ac{UE} trajectory as a temporal sequence of the \acp{BS} that the device traverses (including only successful handover events). 
To align with spatial representations in literature~\cite{yabe2024yjmob100k,park2023fedgeo,esper2024beauty}, we create a squared grid representation with $500m\times500m$ cells ($0.25 km^2$), with \ac{BS} locations in a trajectories mapped to the grid cell they're contained, as seen in \cref{fig:network_grid_example}. 
In this example, we convert a 4-\ac{BS} sequence $[B_{1},B_{51},B_{52},B_{9}]$ into a 3-cell trajectory $[G_{1},G_{5},G_{9}]$, as both $B_{51}$ and $B_{52}$ are in $G_{5}$. 
This also clean trajectories from vertical handovers (\ie change of \ac{RAT} in the same \ac{BS}), which would add points to the trajectory that do not necessarily represent mobility. 
%
%
The final dataset contains 178 unique smartphone models and $53,253$ unique \acp{UE}, which we characterize in \cref{sec:system_analysis}. 
%
With this spatial aggregation, we:
\begin{enumerate}
    \item Normalize the spatial representation by removing biases introduced by irregular base-station coverage and heterogeneous network deployments;
    \item Mitigate spurious mobility signals caused by vertical handovers (\ac{RAT} or Layer changes) and dense horizontal handovers \cite{kalntis2024through};
    \item Reduce spatial sparsity and positioning uncertainty, enabling more stable mobility statistics;
    \item Coarsen spatial precision by avoiding infrastructure-level trajectories, thereby reducing privacy risks from trajectories in ultra-dense deployment environments while preserving large-scale mobility patterns.
\end{enumerate}

\subsection{Mobility indicators}
\label{subsec:mobility_metrics}

We extract three complementary mobility indicators from grid-level trajectories to characterize \acp{UE} patterns.

\paragraph{Entropy.}
We measure trajectory randomness using the temporal-uncorrelated Shannon entropy~\cite{song2010limits}, which captures how evenly a user distributes their time across visited locations. Higher entropy indicates more diverse and less predictable mobility. Following~\cite{song2010limits}, the entropy of a trajectory is defined as:

\begin{equation}
    H = - \sum_{j=1}^{N} p_j \log p_j,
\end{equation}

\noindent where $p_j$ is the proportion of time spent in the $j$-th visited grid cell and $N$ is the number of distinct cells visited in the trajectory. We use the natural logarithm; changing the base only rescales the metric by a constant factor.

\paragraph{\ac{RoG}.}
We quantify traveled distance using the \ac{RoG}, which measures the spatial dispersion of a trajectory around its time-weighted center of mass~\cite{abramowicz1993concept,gonzalez2008understanding}. Intuitively, it captures the typical spatial range of a user's daily movements. Let $\mathbf{l}_j$ denote the centroid of the $j$-th visited grid cell and $t_j$ the dwell time in that cell. The time-weighted center of mass is

\begin{equation}
    \mathbf{l}_{cm} = \frac{\sum_{j=1}^{N} t_j \mathbf{l}_j}{\sum_{j=1}^{N} t_j},
\end{equation}

\noindent and the radius of gyration is then

\begin{equation}
    r_g = \sqrt{\frac{\sum_{j=1}^{N} t_j \left\lVert \mathbf{l}_j - \mathbf{l}_{cm} \right\rVert^2}{\sum_{j=1}^{N} t_j}},
\end{equation}

\noindent where $N$ is the number of visited grid cells.

Entropy and \ac{RoG} are complementary: high entropy with low gyration indicates random movement within a confined area, whereas low entropy with high gyration indicates repetitive trajectories over a large area. To account for the daily and weekly cycles of mobility~\cite{sevtsuk2010does}, we compute both metrics daily for each \ac{UE}.

\cref{fig:example_trajectories} illustrates how these metrics reflect trajectory diversity. User~A (orange) visits more cells over a larger area, which yields higher entropy (4.76) and \ac{RoG} (0.12\,km). User~B (blue) follows a more compact path with fewer movements, which yields lower entropy (1.68) and \ac{RoG} (0.01\,km).

\paragraph{Home location detection.} 
To enable socioeconomic analysis, we approximate the potential area of residence of each \ac{UE} following standard home detection methods~\cite{pappalardo2021evaluation}. We define the home location as the grid cell where each \ac{UE} spends the most time during nighttime hours (00:00–08:00) over the 14-day observation period. This time-constrained approach is widely used in mobility studies, and assumes that individuals are most likely to be at their residence during nighttime~\cite{pappalardo2021evaluation}.

\section{Device Heterogeneity}
\label{sec:system_analysis}

By leveraging the \ac{TAC} of \acp{UE}, we are able to identify 178 smartphone models linked to \acp{UE}. Our next step involves characterizing and exploring the diversity of devices by dividing them into tiers, so we can later connect them to mobility diversity (\cref{sec:mobility_analysis}).

\begin{figure}
    \centering

    \begin{subfigure}[t]{\columnwidth}
        \centering
        \includegraphics[width=\linewidth]{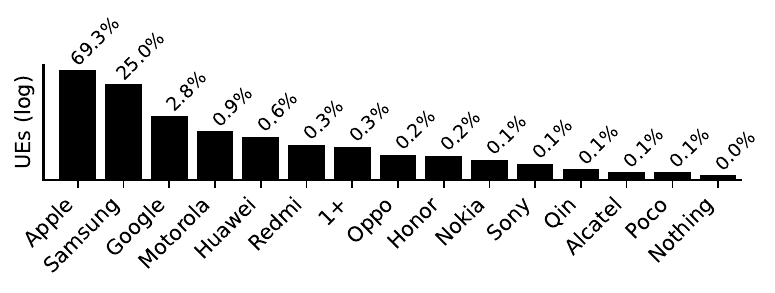}
        \caption{Brand share across \acp{UE}}
        \label{fig:brand_distribution}
    \end{subfigure}
    
    \vspace{0.6em}
    
    \begin{subfigure}[t]{\columnwidth}
        \centering
        \includegraphics[width=\linewidth]{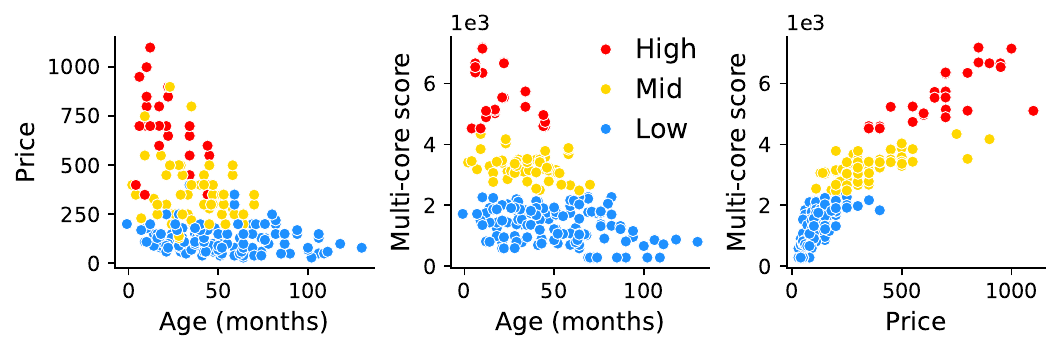}
        \caption{Feature distribution across tiers}
        \label{fig:cluster_results}
    \end{subfigure}
    
    \vspace{0.6em}
    
    \begin{subfigure}[t]{\columnwidth}
        \centering
        \includegraphics[width=\linewidth]{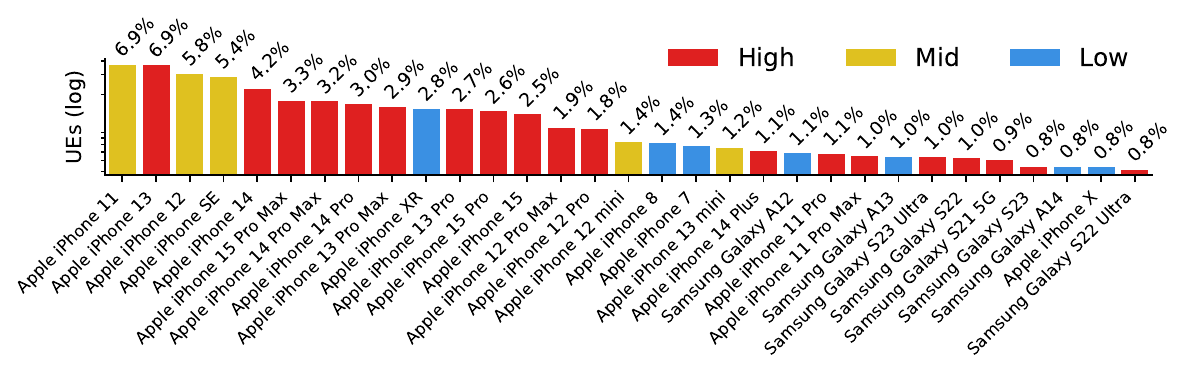}
        \caption{Top 30 smartphone models across \acp{UE}}
        \label{fig:device_distribution}
    \end{subfigure}
    
    \caption{
    Statistics on brands, models and capabilities of smartphone models in our measurements.
    }
    \label{fig:devices}
\end{figure}

\Cref{fig:brand_distribution} presents the share of \acp{UE} across brands, revealing a strong bias toward Apple (69\%) and Samsung (25\%). 
We compare the share of smartphones in the capital versus the rest of the country, using the latest countrywide device catalog available from the \ac{MNO} (collected in 2022), and validate that brand distribution remains essentially the same (Appendix \cref{fig:brand_distribution_2022}). 

To understand if mobility patterns are related to differences in smartphone tiers, we first need to characterize the capabilities of the smartphones present in our data. 
For this, we apply K-means clustering on three smartphone-level features: age in months, estimated price, and Geekbench 6 multi-core performance\footnote{Multi-core scores showed strong correlation with GPU benchmarks and had more entries for entry-level devices, so we opt for it. Data obtained from \url{https://browser.geekbench.com/android-benchmarks} and \url{https://browser.geekbench.com/ios-benchmarks}.}, all referenced to Jul. 2024 (when the data was collected).
By looking for the best value before a sharp decline on the Silhouette score~\cite{rousseeuw1987silhouettes} (Appendix \Cref{sec:silhouette}), we identify 3 as the ideal number of clusters, and define them as: 99 Low-tier, 38 Mid-tier, and 41 High-tier smartphones. As seen on \Cref{fig:cluster_results}, higher-end devices are characterized by being newer, having a higher price and improved computing scores. Cluster centroids of all features (age/price/computing) show how these values reduce towards lower tiers: High (21/689/ 5619), Mid (37/344/3248), and Low (51/117/1376). We use these tiers throughout the work as a proxy of \ac{UE} capability. 

The distribution of \acp{UE} across smartphone tiers is skewed towards High-tier devices (48.5\%), followed by Mid (29.4\%) and Low (22.1\%) tiers; this share remains stable over the studied period. \Cref{fig:device_distribution} highlights this stratification for the top 30 most popular devices: older iPhone models remain prevalent, while premium Apple models (Pro/Pro Max) and Samsung’s flagship S-series dominate, followed by budget Samsung A-series devices. This distribution indicates a bias towards High-tier \acp{UE} in the observed population, consistent with the maturity of European smartphone markets and the strong presence of premium devices~\cite{gsma_smartphone_duopoly_2025}.

We repeat the smartphone clustering with the 2022 countrywide device catalog\footnote{We chose to re-do the cluster with only devices released in 2022 for the countrywide catalog, as our cluster has a bias towards devices' age.} to understand if observed tiers in the capital can be generalized across the country.
We observe contrasting patterns: Mid (43.2\%) and Low (40.6\%) tiers prevail, with High-tier \ac{UE} only at 16.2\%. Cluster centroids (age/price/computing) are: High (37/571/4999), Mid (46/307/3107), and Low (60/111/1279).
Despite similar brand shares, this shift stems from fewer premium Apple models and more legacy devices (iPhone 7/8/XR), as well as more entry-level Samsung models, potentially reflecting capital vs. national wealth disparities (see Appendix Fig. \ref{fig:device_distribution_2022}).
This variation in tiers between capital and countrywide areas was previously observed in studies of 5G adoption rates, where urban areas had a higher adoption than suburban and rural areas, which could be linked to higher-tier smartphones offering 5G earlier and being widely available in urban centers~\cite{mishra2024characterizing}, while other areas of the country might hold onto legacy devices for longer.

This comparison highlights an important limitation of our study and a broader consideration when working with smartphone tiers. The composition of an MNO’s device catalog continuously evolves, as new smartphones are released and users replace their devices, as well as present heterogeneity over space. As a result, device distributions may shift based on the spatiotemporal selection, which can affect cross-year comparisons and the stability of tier definitions. Nevertheless, the presence of Low-end devices and the overall heterogeneity of the device ecosystem remain persistent characteristics that \ac{FL} system designers must consider when building training pipelines. This motivates the periodic re-evaluation of device populations and tier definitions to ensure that analyses and system designs remain aligned with current market conditions.

\textbf{Takeaway.}
We notice a skew towards High-tier devices (High: 48.5\%; Mid: 29.4\%; Low: 22.1\%) in our data. In contrast, nationwide shares show a skewing towards Mid and Low-tier devices (High: 16.2\%; Mid: 43.2\%; Low: 40.6\%). 
%
%
This exposes challenges for \ac{FL} systems, as client selection techniques that do not take into account this potential heterogeneity in the users base can lead to issues during training~\cite{yang2021characterizing}. 
%


\section{Mobility across Device tiers}
\label{sec:mobility_analysis}
Building on the smartphone categorization, we next examine how smartphone tiers correlate with mobility patterns.

\subsection{Device tier and mobility correlation}
\label{subsec:ue_mobility}

For each \ac{UE}, we calculate its daily Entropy and \ac{RoG}, and then calculate the median mobility for each smartphone model. 
We first note that the CDF of both mobility indicators across the smartphone tiers have similar shapes but distinct values across tiers. Performing a two-sided \ac{KS} test across CDF pairs validates that the CDF across tiers are statistically distinct\footnote{$p<0.05$, rejecting the null hypothesis that the CDFs are equal.}.
To quantify this trend, we fit parametric distributions to the mobility indicators on each tier. We fit a Normal distribution for Entropy (fixing $\sigma = 1.05$) on each tier and obtain [$\mu_{\text{High}}=1.742$, $\mu_{\text{Mid}}=1.568$, $\mu_{\text{Low}}=1.437$]\footnote{Entropy R2 scores: High =  0.999, Mid = 0.999, Low = 0.999.}. 
Similarly, we fit Gamma distributions for \ac{RoG} (fixing $\beta=14.3$) on each tier and obtain shape parameters [$a_{\text{High}}=0.758$, $a_{\text{Mid}}=0.655$, $a_{\text{Low}}=0.559$]\footnote{RoG R2 scores: High =  0.994, Mid = 0.992, Low = 0.989.}.
This confirms for us that while Entropy and \ac{RoG} each follow the same distribution across tiers, expected values are different, with High-tier \acp{UE} expecting higher mobility values than Mid and Low tiers for both indicators. 

\begin{figure}
    \centering
    \begin{subfigure}[t]{0.49\columnwidth}
        \centering
        \includegraphics[height=0.42\columnwidth]{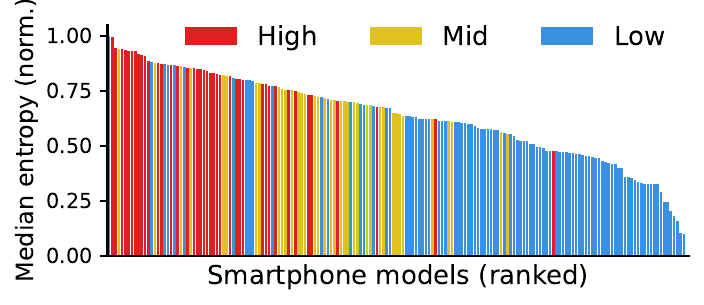}
        \caption{Entropy}
        \label{fig:entropy_devices}
    \end{subfigure}
    \hfill
    \begin{subfigure}[t]{0.49\columnwidth}
        \centering
        \includegraphics[height=0.42\columnwidth]{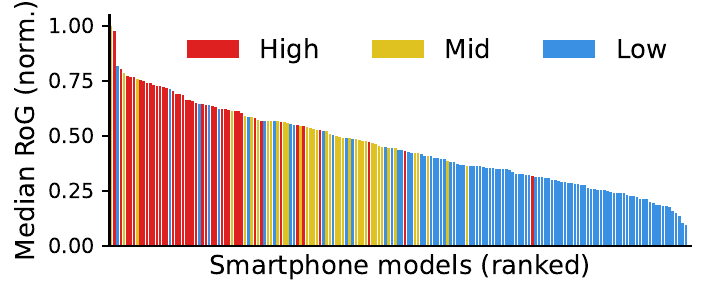}
        \caption{RoG}
        \label{fig:rog_devices}
    \end{subfigure}
    
    \caption{
    Ranking of normalized median mobility across device tiers. We observe a direct correlation between device tier and mobility rankings. 
    }
    \label{fig:mobility_tiers}
\end{figure}

Next, we explore the ranking of the normalized median Entropy (\Cref{fig:entropy_devices}) and \ac{RoG} (\Cref{fig:rog_devices}) across smartphone models. 
We observe a clear tier stratification: High-tier smartphones dominate upper mobility ranks while Low-tier ones appear at the end, matching what was previously observed with the coefficients of the fitted CDFs. 
To validate the relationship between device characteristics and mobility, we create a simple tier-level score by multiplying features from \cref{fig:cluster_results} and calculating the Spearman's rank correlation between smartphone capability and mobility rankings. Results indicate a significant correlation between smartphone tiers and mobility for Entropy ($\rho=0.762$, $p<0.001$) and \ac{RoG} ($\rho=0.770$, $p<0.001$), confirming the observed stratification.

\textbf{Takeaway.} 
Smartphone tiers correlate with human mobility indicators (($\rho=0.762-0.770$, $p<0.001$)).
This correlation is critical for \ac{FL} mobility systems: if devices are grouped into silos by geographic location or temporal availability without accounting for tier-driven mobility differences, the resulting silos will have implicitly imbalanced data distributions, leading to system-induced bias~\cite{maeng2022towards}. Practitioners designing \ac{FL} architectures must consciously account for this coupling.
%
%

\subsection{Weekly mobility patterns}
\label{subsec:mobility_week}

\begin{figure}
    \centering
    
    \begin{subfigure}[t]{0.49\columnwidth}
        \centering
        \includegraphics[width=\linewidth]{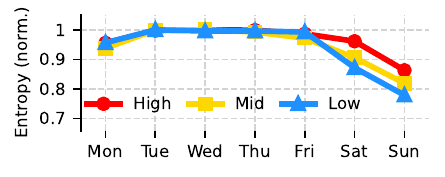}
        \caption{Entropy}
        \label{fig:entropy_ts}
    \end{subfigure}
    \hfill
    \begin{subfigure}[t]{0.49\columnwidth}
        \centering
        \includegraphics[width=\linewidth]{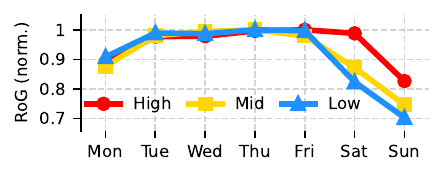}
        \caption{RoG}
        \label{fig:rog_ts}
    \end{subfigure}
    
    \caption{
    Normalized weekly mobility across tiers.
    }
    \label{fig:mobility_ts}
\end{figure}

We next investigate how mobility changes across the week. We calculate median values for each device tier and day of the week, normalizing values by the maximum within each tier to remove volume bias and highlighting temporal patterns. \Cref{fig:mobility_ts} shows a consistent trend for both indicators: peak mobility for each smartphone tier happens from Tuesday to Friday, followed by a decline on the weekend. 

The weekend reduction significantly varies by tier:
High-tier devices have almost no reduction versus week-peak on Saturdays (4\% for Entropy, less than Monday), and their decrease on Sundays is less accentuated (Entropy: 13\%; \ac{RoG} 17\%).
Meanwhile, devices in the Low and Mid tiers see major weekend changes: Entropy and \ac{RoG} see a significant reduction on Saturday (up to 13\% and 17\%, respectively), further increased on Sunday (up to 22\% and 30\%, respectively).  

\textbf{Takeaway.} 
Temporal mobility varies at different levels across tiers. Mid- and Low-tier devices show 13–30\% mobility drops on weekends vs. 4–13\% for High-tier, indicating tier-dependent temporal heterogeneity. The pronounced weekend drops for Mid/Low tiers risk their under-representation during weekends in \ac{FL} systems. Effective client selection must decouple device availability from tier representation.

\subsection{Mobility clusters and smartphone tiers}
\label{subsec:mobility_clusters}

Based on the observed ranking of mobility indicators across tiers (\Cref{fig:mobility_tiers}), we examine next whether smartphones within the same capability tier exhibit similar mobility, or if we also observe variations in mobility within smartphone tiers.
We calculate the distribution of mobility indicators for each model, and use a pairwise \ac{KS} test to quantify similarities, where $D_{n,m} \approx 0$ indicates equality between models $[n,m]$ and $p<0.05$ indicates a statistically significant difference.

\Cref{tab:ks_test_models} shows examples for the difference in Entropy distribution across smartphone models and tiers from the two most popular brands in our data: Samsung Galaxy (SG) models and Apple iPhone (AIp). We observe that the example smartphone models within the same tier had similar Entropy distribution (lower $D_{n,m}$ and $p>0.05$), while models across tiers had dissimilarities in their Entropy (higher $D_{n,m}$ and $p<0.05$)


From the insights of this example, we expand the two-sided \ac{KS} test across all $178$ smartphone models and both indicators, and use these results as a feature for a hierarchical clustering algorithm, with the goal of finding smartphone models with similar mobility statistics. 
The Silhouette score analysis indicates $k=3$ as the ideal number of clusters for both metrics (Appendix \Cref{fig:silhouette_entropy,fig:silhouette_rog}), matching the number of smartphone tiers obtained on \cref{sec:system_analysis}. 
By observing the metric values in each mobility cluster, we categorize them as High/Mid/Low Entropy/RoG. The distribution of smartphone models for (High/Mid/Low) Entropy and \ac{RoG} are (50/72/56) and (57/54/67), respectively.

\begin{figure}
    \centering
    \includegraphics[width=0.8\linewidth]{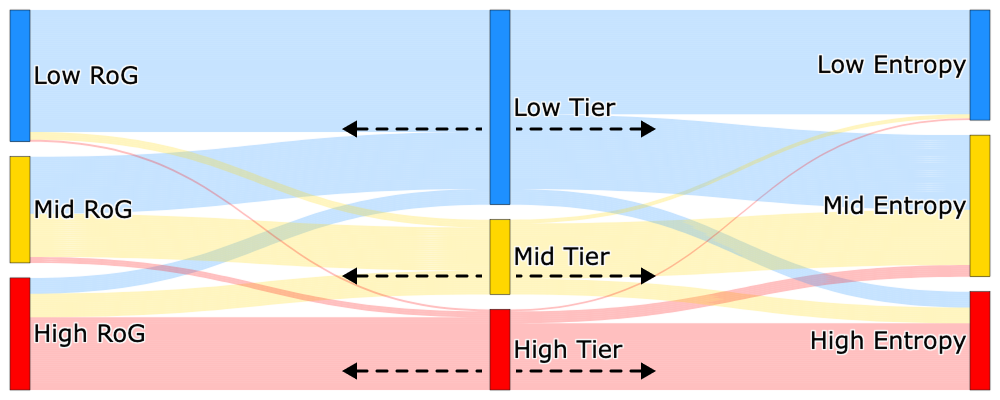}
    \caption{\small Sankey diagram, of how smartphone model tiers (middle) relate to mobility clusters, via the RoG (left) and Entropy (right).
    }
    \label{fig:sankey_clusters}
\end{figure}

We use a Sankey diagram (\Cref{fig:sankey_clusters}) to observe how smartphone models relate between the smartphone tiers (middle), Entropy tiers (right) and \ac{RoG} (left) tiers. Interestingly, the edge clusters (High/Low smartphone tiers) show the strongest similarity across both the Entropy and \ac{RoG} clusters: High/Low mobility clusters are mostly composed of High/Low-tier smartphone models. 
The Mid mobility clusters exhibit more variation, being composed by a portion of Low and most Mid-tier models. This can be attributed to the high dynamicity of the Low tier, which includes both recent entry-level models and older devices that were initially positioned in higher tiers but have depreciated over time.

Indeed, while the Low smartphone tier contains the most models, we note that it splits between both Low and Mid mobility clusters: Low-tier devices that tend to have Mid mobility (containing the iPhone X-series and newer Samsung A-series devices) are newer than the ones at Low mobility (which contains iPhones 8 and older, plus the old Samsung J-line). 
We also note that some devices in the Mid tier belong to High mobility clusters, which could be explained by their "budget flagship" marketing strategies targeting younger demographics. This category includes brands such as Nothing and Poco, as well as the Pixel A-series.

Pappalardo et al. found that the diversity of trajectories collected from signaling data at municipality scale was correlated with users’ socioeconomic status \cite{pappalardo2015using}. Specifically, greater mobility diversification was associated with higher wealth and lower levels of deprivation. In contrast, Song et al. reported that mobility entropy is highly stable across different subpopulations defined by personal characteristics such as gender and age group \cite{song2010limits}. Pappalardo et al. suggest that this apparent discrepancy may stem from differences in the scale of analysis between city-level individual-level behavior.
Our findings indicate that higher-end smartphone models tend to exhibit higher entropy and RoG in their city-level movements, matching previous findings on smartphone ownership being connected to educational level and socioeconomic characteristics~\cite{wenz2023second}. We further explore these patterns in our data over the next Section. 

\begin{table}[htbp]
    \centering
    \small
    \rowcolors{2}{white}{lightgray}
    \caption{Pairwise KS test results across smartphone models for observing Entropy distribution similarity.}
    \begin{tabular}{llllrrl}
        \toprule
        \textbf{Model \textit{n}} & \textbf{Tier} & \textbf{Model \textit{m}} & \textbf{Tier} & \textbf{$D_{n,m}$} & \textbf{\textit{p}} & \textbf{Sim.} \\
        \midrule
          SG S24 U & \textcolor{highred}{High} & AIp 15 PM & \textcolor{highred}{High} & 0.061 & 0.27 & \cmark \\
          SG A54 & \textcolor{midgold}{Mid} & AIp 12 & \textcolor{midgold}{Mid} & 0.054 & 0.32 & \cmark \\
          SG A13 & \textcolor{lowblue}{Low} & SG A22 & \textcolor{lowblue}{Low} & 0.072 & 0.37 & \cmark \\
          SG S24 U & \textcolor{highred}{High} & SG A54 & \textcolor{midgold}{Mid} & 0.147 & 0.00 & \xmark \\
          SG S24 U & \textcolor{highred}{High} & SG A13 & \textcolor{lowblue}{Low} & 0.269 & 0.00 & \xmark \\
          SG A54 & \textcolor{midgold}{Mid} & SG A13 & \textcolor{lowblue}{Low} & 0.146 & 0.00 & \xmark \\
        \bottomrule
    \end{tabular}
    \label{tab:ks_test_models}
\end{table}

\textbf{Takeaway.}  
Mobility clusters derived from Entropy/\ac{RoG} align closely with device tiers, meaning that client selection based on mobility similarity will inherently approximate tier balancing, confirming the interplay between device specifications and their associated mobility. 

\subsection{Devices and mobility vs. Index of Deprivation Rank}
\label{subsec:mobility_income}

Based on the findings from the previous sections, we investigate whether the diversity of smartphone tiers and mobility tiers can be explained by socioeconomic indicators from the city’s official census office. 

We identify the square grid corresponding to each \ac{UE} residence (\Cref{subsec:mobility_metrics}) and associate it with an Index of Deprivation Rank, a composite measure that ranks small areas based on overall socioeconomic disadvantage by combining indicators such as income, employment, health, education, crime, housing, and living environment. It ranks areas by deprivation, with lower ranks indicating more deprived areas (\ie more disadvantaged) and higher ranks indicating less deprived areas (\ie less disadvantaged). This statistic is collected at geographic units covering 1,000 to 3,000 inhabitants, capturing neighborhood-level variation.
We compute deprivation distributions for each device tier and mobility cluster, considering all \acp{UE} with a device model contained in each cluster. We perform a \ac{KS} test to evaluate whether distributions across High/Mid/Low clusters differ, with results reported in \Cref{tab:income_ks_test}. 

For the \textit{device tier}, all comparisons yield p-values $<0.05$, indicating statistically significant differences in deprivation across \ac{UE} in each tier. Although the low \ac{KS} scores suggest limited distributional separation, visual inspection of the CDFs (Appendix \Cref{fig:deprivation_cdfs}) shows that smartphone tier captures an inequality structure: the deprivation distribution for the Low tier is shifted to the left (\ie higher ranking, more deprived), indicating that Low-tier \acp{UE} are more likely to reside in more deprived regions than High- and Mid-tier \acp{UE}, matching previous findings on second-level smartphone divide and ownership~\cite{wenz2023second}

For \textit{mobility tiers}, we observe mostly non-significant p-values across deprivation distributions for Entropy clusters (with the exception of High-Mid). 
In contrast, for \ac{RoG} mobility clusters, we observe statistical significance in the \ac{KS} test and a similar CDF pattern to device tiers: \acp{UE} with lower \ac{RoG} are more likely to come from more deprived regions than those in High and Mid \ac{RoG} clusters, suggesting that more deprived \acp{UE} exhibit more constrained urban mobility and a smaller activity space. 

\textbf{Takeaway.} 
Lower-tier devices are more common in deprived neighborhoods, and these users also have smaller activity spaces (lower RoG). \ac{FL} fairness thus has socioeconomic implications: tier-aware client balancing not only equalizes technical performance but also ensures that this population is not systematically under-represented in trained models, underscoring fairness as a proxy for inclusion of marginalized populations in data-driven systems.


\begin{table}[htbp]
    \centering
    \rowcolors{2}{white}{lightgray}
    \small
    \caption{KS test comparing the Index of Deprivation Rank distributions across populations in each tier. Values report $D$ with $p$-values in parentheses. $*$ indicates statistical significance ($p < 0.05$).}
    \label{tab:income_ks_test}
    \setlength{\tabcolsep}{4pt}
    \begin{tabular}{lccc}
        \toprule
         & \textbf{High--Mid} & \textbf{High--Low} & \textbf{Mid--Low} \\
        \midrule
        \textbf{Device tier} & 0.03 (<0.001$^*$) & 0.03 (<0.001$^*$) & 0.05 (<0.001$^*$) \\
        \textbf{Entropy tier}     & 0.01 (0.029$^*$)  & 0.01 (0.776)      & 0.01 (0.210)      \\
        \textbf{RoG tier}         & 0.01 (<0.001$^*$) & 0.05 (<0.001$^*$) & 0.06 (<0.001$^*$) \\
        \bottomrule
    \end{tabular}
\end{table}

\section{Evaluating heterogeneity in FL Next-Location Prediction}
\label{sec:eval}


Earlier, we identified smartphone-tier imbalance and its link to mobility heterogeneity. We now evaluate its downstream effect on federated next-location prediction, a privacy preserving setting suitable for sensitive trajectory data~\cite{belal2024survey, gecer2024federated, feng2020pmf, park2023fedgeo}. Specifically, we ask: (\textit{Q1}) does natural tier imbalance harm inter-tier fairness under standard \ac{FL}? (\textit{Q2}) can tier-aware client selection improve fairness without accuracy loss? and (\textit{Q3}) if tier metadata is unavailable, can FedProx still mitigate fairness gaps? This section structures our evaluation methodology to explore these questions.

\subsection{Data preparation}
\label{sec:data_prep}

Leveraging the trajectories generated from mobile network handovers (\Cref{subsec:trajectories_processing}), we perform the following processing pipeline for each trajectory to prepare them for our experiments. This pre-processing follows standard practice in federated mobility prediction, where trajectories must be converted into fixed-length, model-ready sequences while preserving their temporal and spatial structure~\cite{feng2020pmf,park2023fedgeo}: 

\begin{enumerate}
    \item We first apply z-score standardization to the latitude and longitude pairs, resulting in variables with zero mean and unit standard deviation. This ensures that both spatial axes are on a comparable scale and improves the suitability of the data for \ac{ML} models.

    \item Trajectories are represented as sequences of handovers, resulting in a variable number of points across \acp{UE} for a given day. Since \ac{ML} models typically require fixed-length inputs, we partition trajectories into non-overlapping subsequences of length $\mathbf{N}=20$. Sequences shorter than $\mathbf{N}$ are discarded to ensure consistent input dimensionality. For example, a trajectory with 63 points is split into three independent chunks of 20 points each, with the remaining 3 points discarded. The mean number of points across trajectories per tier is High (155), Mid (121), and Low (99).
    
    \item We augment the data with contextual information about each trajectory, including:
    \textit{1)} day of the week;
    \textit{2)} inter-event time differences between consecutive points $(p, p-1)$, capturing variations in temporal spacing and providing a proxy for velocity (as handover events are not evenly spaced in time);
    \textit{3)} time of arrival at point $p$, encoded as a cyclical feature using sine and cosine~\cite{feng2018deepmove} (Appendix \Cref{subsec:time_encoding});
    \textit{4)} distance traveled between consecutive points $(p, p-1)$, computed using the Haversine formula (Appendix \Cref{subsec:haversine}).
    
    \item Therefore, each data sample consists of a trajectory chunk of length $N=20$, spanning points $[p-N+1, \dots, p]$, along with the day of the week and aligned arrays of time of arrival, inter-event time differences, and traveled distances. These features are used as input to predict the next location $p+1$.

\end{enumerate}

We perform a temporal train/validation/test split. Since mobility patterns differ between weekdays and weekends (\Cref{fig:mobility_ts}), the training set must include at least one full week of observations to capture these temporal variations. Accordingly, we use trajectories from Jul.~22 to Jul.~30,~2024 for training, and Jul.~31,~2024 (Wednesday) for validation, as weekday mobility patterns are relatively stable across days. 
Given that the dataset spans only two weeks, allocating a full week for testing would leave insufficient data for training. Instead, we define the test set as Aug.~1 to Aug.~4,~2024 (Thursday--Sunday), which includes both weekday and weekend periods to evaluate model generalization.

\begin{figure}
    \centering
    
    \begin{subfigure}[t]{0.49\linewidth}
        \centering
        \includegraphics[width=\linewidth]{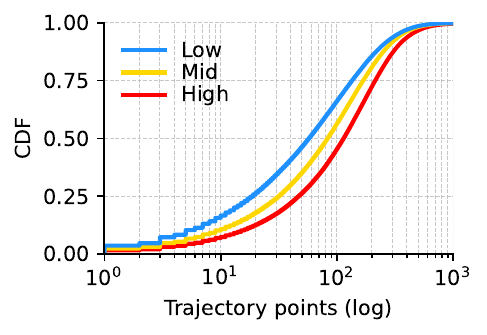}
        \caption{}
        \label{fig:trajectory_points_cdf}
    \end{subfigure}
    \hfill
    \begin{subfigure}[t]{0.49\linewidth}
        \centering
        \includegraphics[width=\linewidth]{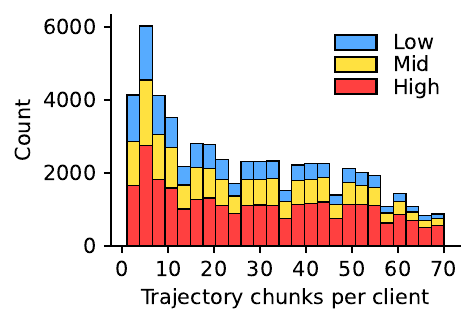}
        \caption{}
        \label{fig:row_distribution_across_tiers}
    \end{subfigure}
    
    \caption{
    Across tiers:
    (a) Distribution of number of points per trajectory
    (b) Distribution of trajectory chunks per client. If unaccounted, this leads to fairness issues due to different training set sizes.
    }
    \label{fig:trajectory_sizes}
\end{figure}

After applying the processing pipeline, we identify an additional source of imbalance that may further affect fairness in federated training: \acp{UE} in the High smartphone tier tend to generate daily trajectories with more points (\Cref{fig:trajectory_points_cdf}). As trajectories are partitioned into fixed-length chunks, \acp{UE} in the High tier consequently produce more trajectory chunks (\ie more data) than those in the Mid and Low tiers (\Cref{fig:row_distribution_across_tiers}). While clients with a small number of chunks are relatively balanced across tiers, those generating a large number of chunks ($>50$) are predominantly from the High tier. If smartphone tier is not accounted for when composing silos, the natural skew toward High-tier devices (\Cref{sec:system_analysis}) will be reflected in each silo's client composition. Combined with High-tier clients generating more trajectory chunks, each silo receives disproportionately more training signal from High-tier users — amplifying tier imbalance within silos and potentially degrading fairness across tiers. These same dynamics would affect a cross-device deployment as well, where the natural skew would cause High-tier devices to dominate per-round client sampling, while Low-tier devices risk being deprioritized or systematically excluded due to straggler-avoidance policies~\cite{yang2021characterizing, yang2018applied}.

\subsection{Next-location prediction model}
\label{sec:ml_framework}


To ensure comparability with prior work, we follow the standard formulation used in federated mobility prediction rather than introducing a novel architecture or loss function.
Next, we describe the next-location prediction problem and how we approach it in a federated manner. We formulate it as a regression task, where the model predicts the latitude and longitude coordinates of the next grid cell given a trajectory chunk and contextual features. 

\subsubsection{Model architecture}

We employ a vanilla \ac{RNN} architecture, which is well suited for capturing temporal dependencies in sequential trajectory data~\cite{liu2016predicting,feng2020pmf, zhang2023federated}, and performed best in our centralized evaluation across model architectures (Appendix \Cref{subsec:model_loss_eval}). The model takes as input the seven features described in \Cref{sec:data_prep} and consists of a single RNN layer with 64 hidden units and \texttt{tanh} activation, followed by a fully connected output layer that maps the final hidden state to the coordinate pair of the predicted next-location. 
Trajectory chunks of length $\mathbf{N}=20$ are processed using \texttt{batch\_first=True}, and the hidden state at the final time step is used to predict the next location~\cite{feng2020pmf, zhang2023federated}.


\subsubsection{Loss function}

Due to the spatial nature of our prediction task, we use the Haversine distance as both our training loss and evaluation metric (formulation in Appendix \Cref{subsec:haversine}). Haversine accounts for the Earth's curvature and yields errors directly interpretable in meters, making it the standard choice for location prediction tasks in the federated mobility literature~\cite{feng2020pmf,zhang2023federated}. Predictions are made in z-score normalized space and then inverse-transformed back to latitude/longitude coordinates before computing the loss, ensuring the model optimizes directly for real-world distance error.

\subsection{Federated learning architecture}
\label{sec:fl_framework}

We adopt a \textit{cross-silo} \ac{FL} architecture with 3 silos, each representing a collection of clients (\acp{UE}) that train local models on their trajectory data before aggregating updates at a central server.

\subsubsection{Motivation for cross-silo FL}
Our choice of cross-silo over cross-device \ac{FL} is motivated by three considerations:

\noindent \textit{Computational efficiency:} Cross-silo \ac{FL} allows us to group devices with similar characteristics into silos, where each silo can perform local aggregation before communicating with the central server. This reduces communication rounds and leverages the correlation between device tiers and mobility patterns we established in \Cref{sec:mobility_analysis}. By using device tier metadata (obtainable from \ac{IMEI} without processing trajectories), we can organize silos without the computational overhead of mobility-based clustering~\cite{zhang2023federated,esper2024beauty} or entropy-based client selection~\cite{park2023fedgeo}.

\noindent \textit{Privacy preservation:} Grouping devices by hardware characteristics (rather than analyzing individual mobility patterns to form clusters) reduces the need for privacy-invasive trajectory pre-processing. Device metadata is already available to \ac{MNO}s through standard network operations, whereas mobility-based grouping requires analyzing user trajectories before training begins~\cite{park2023fedgeo, zhang2023federated}.

\noindent \textit{Practical deployability:} The cross-silo architecture aligns with a realistic \ac{MNO} infrastructure and \ac{FL} scenario, where an operator partitions its subscriber base across organizational or geographic network segments (\eg regional data centers or network slices) rather than coordinating thousands of individual devices. In this deployment model, each silo acts as a network-managed aggregation point, aligning with how \ac{MNO} already segment their infrastructure for traffic engineering and QoS management, as well as how the data from this work is currently collected (\Cref{subsec:network_infra}). While cross-device \ac{FL} would reflect a fully decentralized deployment where each \ac{UE} participates independently, such scenarios require assumptions about \ac{UE} availability, battery, and connectivity that go beyond the scope of this study. The cross-silo abstraction is therefore a deliberate modeling choice, isolating the effect of device-tier-induced data heterogeneity on federated training fairness, which is the central research question.

\subsubsection{Silo configuration.}
We experiment with multiple silo configurations to evaluate how the distribution of device tiers across silos impacts model fairness and accuracy. Each silo is assigned a specific tier composition (detailed in \Cref{sec:experiment_desc}) and maintains: (i) a local dataset comprising trajectories from its assigned clients; (ii) a local copy of the global model, initialized with the same weights; (iii) local training state (optimizer parameters, learning rate schedule).
\subsubsection{Training protocol.}
We follow a standard \ac{FL} training protocol, running $T=10$ rounds of model aggregation. In each round, every silo performs $E=10$ local epochs using mini-batch gradient descent with a batch size of 128. We use the Adam optimizer with learning rate $\eta=0.001$ (without momentum). The data are split into training, validation, and test sets as described in \Cref{sec:data_prep}. 

We evaluate two model aggregation strategies. The first is \textbf{\ac{FedAvg}}~\cite{mcmahan2017communication}, which performs weighted averaging based on silo dataset sizes:

\begin{equation}
    w_{t+1} = \sum_{k=1}^{K} \frac{n_k}{n} w_k^t
\end{equation}

\noindent where $n_k$ denotes the number of samples in silo $k$, $n=\sum_k n_k$, and $w_k^t$ represents the local model parameters after round $t$.

The second is \textbf{\ac{FedProx}}~\cite{li2020federated}, which introduces a proximal term in the local objective to address data heterogeneity:

\begin{equation}
    \min_{w} \mathcal{L}_k(w) + \frac{\mu}{2}\|w - w^t\|^2
\end{equation}

\noindent where $\mu=0.001$ is the proximal coefficient and $w^t$ denotes the global model at round $t$. This regularization term penalizes large deviations from the global model during local training, thereby improving stability under heterogeneous data.

Both methods are evaluated under the same training schedule to ensure that differences in performance arise from the aggregation strategy rather than from optimization settings.
\subsubsection{Implementation.}
We implement our \ac{FL} experiments using Flower~\cite{beutel2020flower}, a flexible \ac{FL} framework that supports custom aggregation strategies and heterogeneous client configurations. Models are implemented in PyTorch\footnote{\url{https://pytorch.org}} and trained on NVIDIA GPUs with CUDA support. Each silo’s local training runs as a separate Flower client process. We use Flower’s simulation backend to orchestrate a 3-silo federation, allocating dedicated GPU resources per silo (1 GPU and 2 CPU cores each). 



\subsection{Fairness and accuracy evaluation}
\label{sec:fairness_evaluation}

We evaluate silo configurations along two dimensions: accuracy, measured via Haversine distance, and fairness across device tiers, as aggregate accuracy can appear strong even when specific tiers are systematically disadvantaged. To assess fairness, we compare the full error distributions across tiers using statistical tests~\cite{taskesen2021statistical}, rather than relying on aggregated statistics alone. This follows the \textit{separation} criterion in group fairness~\cite{barocas2023fairness, hardt2016equality}, also referred to as error-rate parity, which requires prediction errors to be comparable across groups, and extends naturally from classification to regression by comparing error distributions rather than error rates.

Each experiment (\Cref{sec:experiment_desc}) produces a distinct error distribution for every device tier, and our objective is to quantify the divergence between these distributions, where large discrepancies across tiers indicate higher levels of unfairness. 
We initially explored ANOVA to compare mean errors and assessed its assumptions through diagnostic tests. Although Levene’s test indicated heteroscedasticity, which would favour Welch’s ANOVA test, the error distributions were strictly positive, strongly right-skewed, and far from normal. These violations render parametric approaches inappropriate; therefore, we rely on non-parametric methods.

We therefore employ the \textit{Kruskal--Wallis test}~\cite{kruskal1952use}, which evaluates whether significant differences exist between three or more independent groups by comparing the ranks of their observations. The test produces the $H$ statistic, where larger values indicate stronger differences between groups, \ie a proxy to tell us if unfairness exists. To quantify the magnitude of unfairness, we compute the \textit{effect size} ($\epsilon^{2}$)~\cite{tomczak2014need}, which ranges from 0 to 1, with lower $\epsilon^{2}$ values preferred as they indicate that device tier explains little of the variance in prediction error. Therefore, we will use the $\epsilon^{2}$ values as our proxy of fairness improvement $\Delta_{\text{Fairness}}$ across experiments~\cite{hardt2016equality,borkan2019nuanced,barocas2023fairness}.


When the Kruskal--Wallis test indicates statistically significant differences across all tiers, we perform a Dunn post-hoc test~\cite{dunn1964multiple} to obtain the pairwise differences across tiers (High--Low, High--Mid, and Low--Mid). We convert the raw p-values from Dunn's test to z-scores via the inverse normal CDF to obtain a continuous measure of pairwise disparity magnitude. Larger $z$-scores from this test indicate stronger discrepancies and higher levels of unfairness, while smaller values indicate more similar error distributions between pairs.

Finally, by comparing with the baseline experiment and report relative improvements in both fairness across all tiers ($\Delta_{\text{Fairness}}$), fairness between pairwise tiers ($\Delta_{High x Low}$, $\Delta_{High x Mid}$, $\Delta_{Low x Mid}$) and accuracy ($\Delta_{\mu}$ and $\Delta_{Median}$) as:

\begin{equation}
    \Delta_{base,exp} = \frac{T_{base} - T_{exp}}{T_{base}} \times 100,
\end{equation}

\noindent where $T_{base}$ and $T_{exp}$ denote the value of a given metric for the baseline and the evaluated experiment, respectively.

\subsection{Experiments description}
\label{sec:experiment_desc}

\begin{figure}
    \centering
    
    \begin{subfigure}[t]{0.48\linewidth}
        \centering
        \includegraphics[width=\linewidth]{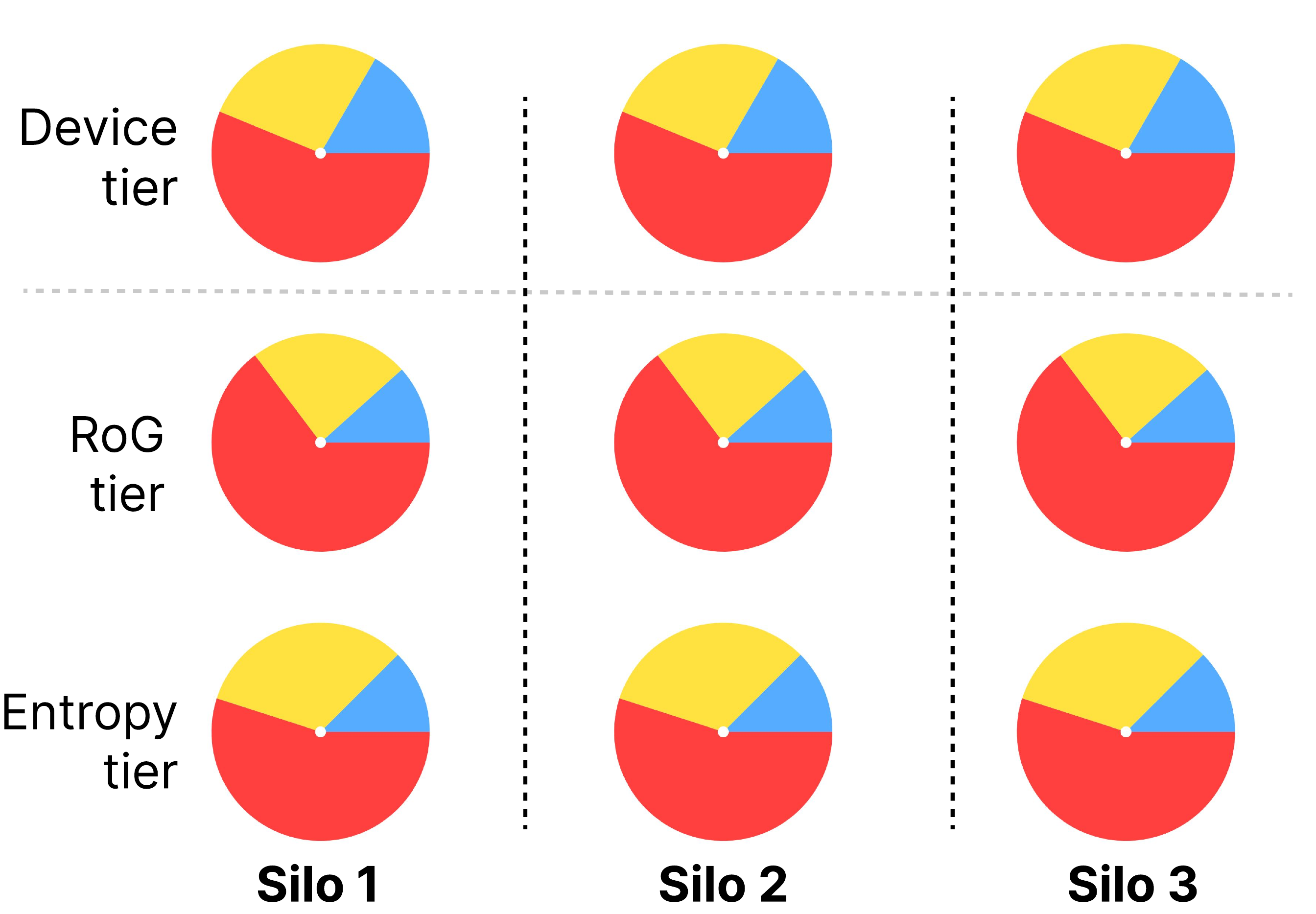}
        \caption{Natural tier distribution}
        \label{fig:silo_natural}
    \end{subfigure}
    \hfill
    \begin{subfigure}[t]{0.48\linewidth}
        \centering
        \includegraphics[width=\linewidth]{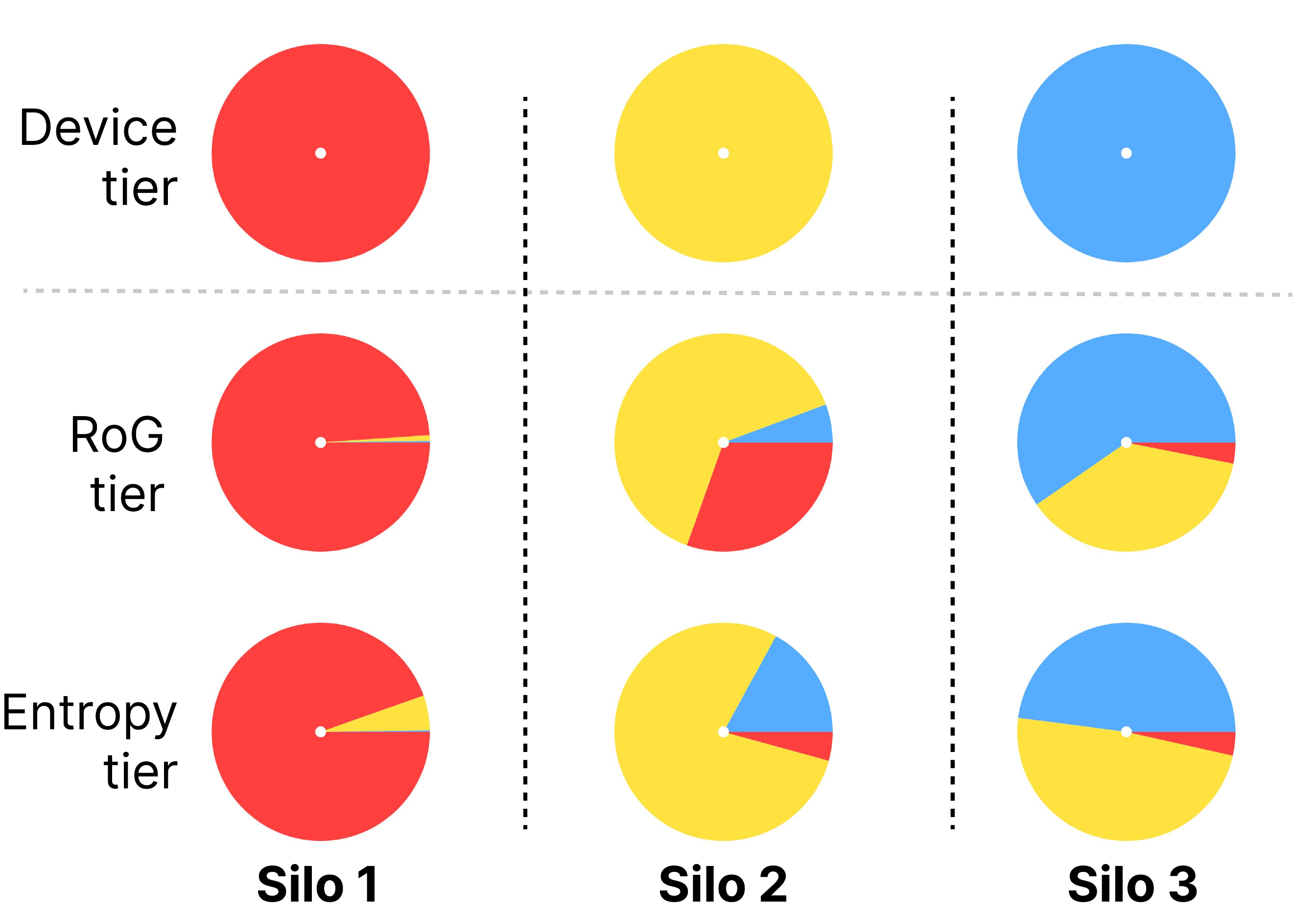}
        \caption{Fully tiered}
        \label{fig:silo_fully}
    \end{subfigure}

    \vspace{0.5em} 

    \begin{subfigure}[t]{0.48\linewidth}
        \centering
        \includegraphics[width=\linewidth]{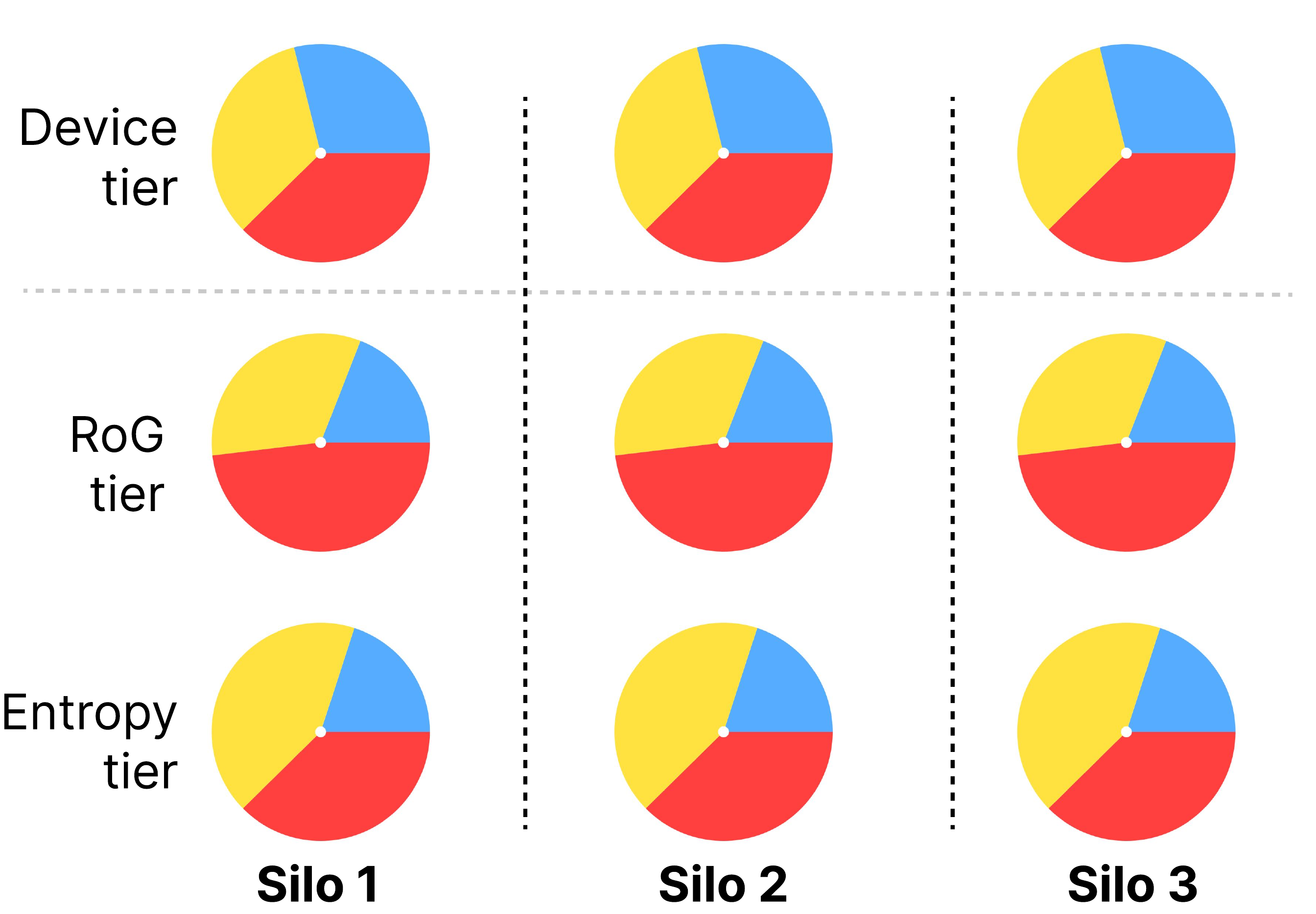}
        \caption{Balanced device tiers}
        \label{fig:silo_balanced}
    \end{subfigure}
    \hfill
    \begin{subfigure}[t]{0.48\linewidth}
        \centering
        \includegraphics[width=\linewidth]{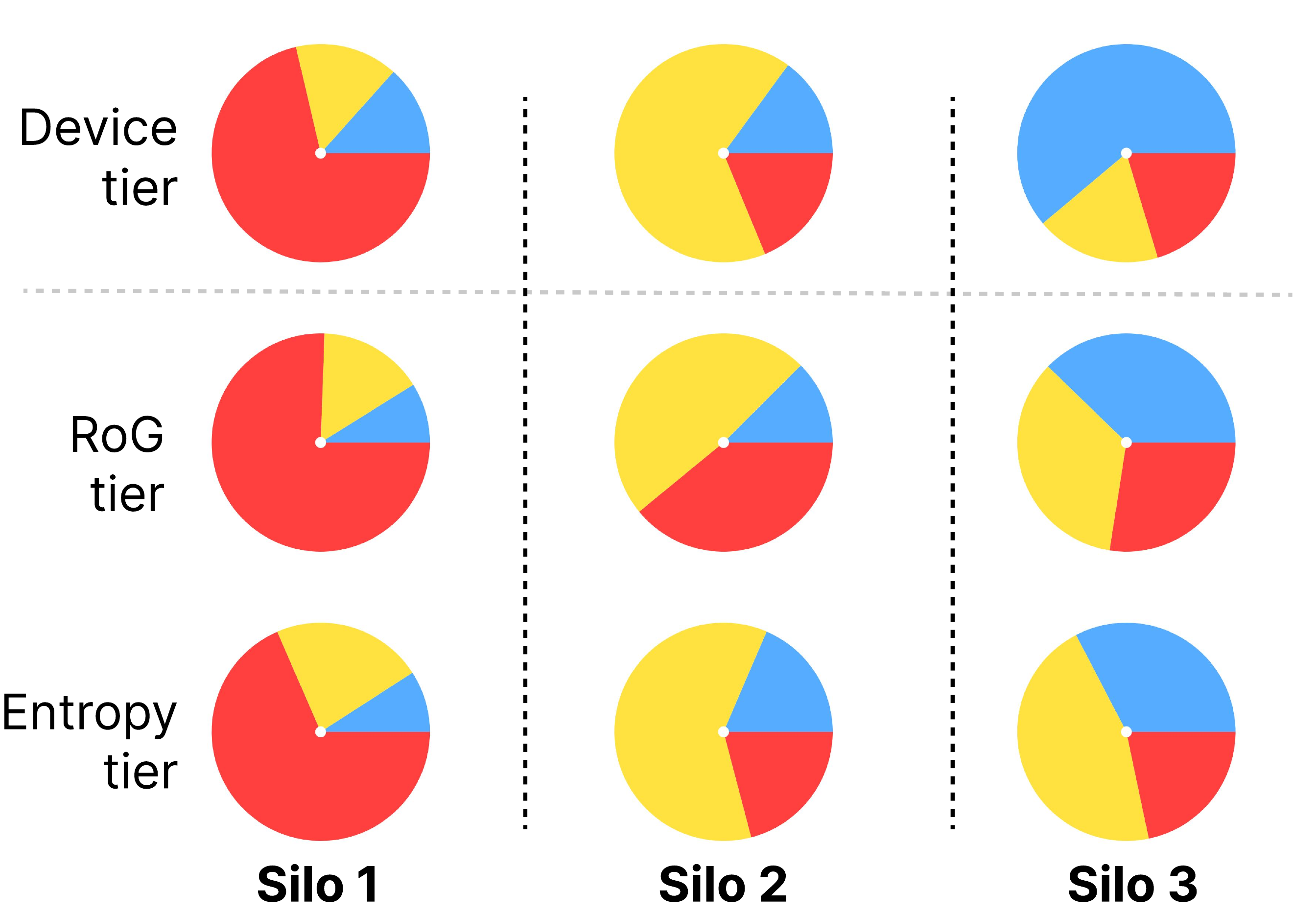}
        \caption{Mildly tiered}
        \label{fig:silo_mildly}
    \end{subfigure}

    \caption{Client selection by device tier and its effect on the distribution of \acp{UE} across device, entropy, and RoG tiers under different silo strategies.
    }
    \label{fig:silo_structure}
\end{figure}

\subsubsection{Silo configurations.}
We explore how device-induced mobility data heterogeneity affects model fairness and accuracy. We argue that this can arise as a by-product of client selection techniques, where an imbalance in selected smartphones leads to data imbalance and, consequently, potential fairness and accuracy issues.

\Cref{fig:silo_structure} illustrates possible smartphone tier distributions across silos under different client selection strategies, along with the resulting distribution of \ac{RoG} and Entropy tiers.
\Cref{fig:silo_natural} represents the \textit{natural distribution} when smartphone tiers are ignored and clients are randomly selected. This leads to imbalanced tiers across silos, following the pattern observed in \Cref{sec:system_analysis}: an over-selection of devices with High \ac{RoG} and Entropy, driven by the population skew toward High-tier smartphones.
The second case (\Cref{fig:silo_balanced}) acknowledges this imbalance and attempts to address it by \textit{balancing device selection} across silos. In this setting, all three tiers are equally represented, resulting in a more balanced distribution of mobility tiers. 

Next, we consider \textit{unbalanced} silos, \ie when tier distributions differ across silos. This may occur in a system-driven manner, such as assigning silos that are \textit{fully dedicated to each tier} (\Cref{fig:silo_fully}), which can benefit FL next-location prediction under mobility heterogeneity~\cite{esper2024beauty}.
Another scenario arises when devices are sampled from different spatial locations with \textit{varying smartphone tier distributions} (\Cref{fig:silo_mildly}), leading to imbalance both within and across silos. These selection strategies further induce imbalance in mobility tiers. The fully tiered approach (\Cref{fig:silo_fully}) shows that mobility silos are not completely separated even when each silo contains a single device tier. This occurs because device tiers do not map perfectly to mobility levels (\Cref{fig:sankey_clusters}).

\subsubsection{Experiments.}
Based on the silo configurations (\Cref{fig:silo_structure}), we describe in \Cref{tab:evaluation_scenarios} the experiments where we explore the impact of device and data heterogeneity on our \ac{FL} system, and how silo strategies can address issues in fairness and accuracy. 
We consider as the baseline scenario the natural smartphone tier distribution across silos using FedAvg, and train models for each silo configuration using both FedAvg and FedProx.
We also evaluate the impact of excluding Low-tier devices from silo composition, a common policy of tier-exclusion for addressing stragglers and their impact in \ac{FL}~\cite{yang2021characterizing, yang2018applied}. 
Our main goal is to understand if, by acknowledging the imbalance of tiers, which measure taken either in client selection or model aggregation techniques helps the most in improving the fairness of the model, without degrading the accuracy of predictions. 

\begin{table}[htbp]
    \centering
    \rowcolors{2}{white}{lightgray}
    \small
    \caption{Description of silo evaluation scenarios. All scenarios are evaluated with FedAvg and FedProx for model aggregation.}
    \label{tab:evaluation_scenarios}
    \setlength{\tabcolsep}{4pt}
    \begin{tabular}{p{2.3cm} p{4.7cm}}
        \toprule
        \textbf{Scenario} & \textbf{Description} \\
        \midrule
        Baseline / Natural & Natural distribution of the data. \\
        Exclude Low        & Low-tier devices are excluded. \\
        Balanced           & All silos have equal tier proportions (33/33/33). \\
        Mildly Tiered      & Each silo is biased toward one tier: High (70/15/15), Mid, or Low. \\
        Fully Tiered       & Each silo contains only one tier (High, Mid, or Low). \\
        \bottomrule
    \end{tabular}
\end{table}

\section{Evaluation results and discussion}
\label{sec:results}

\begin{table*}[t]
    \caption{Fairness and accuracy across experiments and improvements against baseline (Natural distribution with FedAvg).}
    \centering
    \rowcolors{2}{white}{lightgray}
    \begin{tabular}{lllllllll}
        \toprule
        \textbf{Scenario} & \textbf{Agg.} & \textbf{$\mu$ (m)} & \textbf{$\sigma$} (m) & \textbf{Median (m)} & \textbf{90th (m)} & \textbf{$\Delta_{\mu}$} & \textbf{$\Delta_{Median}$} & \textbf{$\Delta_{Fairness}$} \\
        \hline
        Natural      & FedAvg     & 737.31     & 1085.51     & 308.38     & 1867.04     & \textit{Baseline} & \textit{Baseline} & \textit{Baseline} \\
        Exclude Low   & FedAvg     & 737.68     & 1083.11     & 314.15     & 1856.62     & -0.05\% & -1.87\%      & -16.23\%           \\
        Balanced      & FedAvg     & 730.65     & 1055.39     & 313.30     & 1845.06     & 0.90\%  & -1.60\%     & \textbf{34.49\%}   \\
        Mildly tiered & FedAvg     & 741.97     & 1086.99     & 321.80     & 1848.17     & -0.63\% & -4.35\%     & -48.32\%           \\
        Fully tiered  & FedAvg     & 752.89     & 1078.85     & 337.90     & 1861.88     & -2.11\% & -9.57\%     & -36.99\%           \\
        \hline
        Natural       & FedProx    & 729.58     & 1074.17     & 298.56     & 1869.42     & 1.05\% & \textbf{3.19\%}       & 21.46\%            \\
        Exclude Low   & FedProx    & 737.14     & 1078.23     & 316.57     & 1855.90     & 0.02\% & -2.66\%      & 4.04\%             \\
        Balanced      & FedProx    & 725.82     & 1055.10     & 312.98     & 1839.76     & \textbf{1.56\%} & -1.49\% & 19.94\%         \\
        Mildly tiered & FedProx    & 741.46     & 1088.30     & 319.04     & 1853.66     & -0.56\% & -3.46\%	     & -20.58\%           \\
        Fully tiered  & FedProx    & 750.31     & 1090.08     & 326.58     & 1862.43     & -1.76\%  & -5.90\%	    & -0.99\%            \\
        \hline
        Central & - & 737.24 & 1108.17 & 273.75 & 1922.24 & 0.01 \% & 11.22\%  & -6.87 \% \\
        \bottomrule
    \end{tabular}
    \label{tab:agg_results}
\end{table*}

\begin{table*}[ht]
    \centering
    \caption{Dunn post-hoc z-scores for model inter-tier fairness (High vs Low, High vs Mid, Low vs Mid). Lower z-scores indicate a smaller accuracy delta between tiers.}
    \label{tab:fairness-dunn}
    \rowcolors{2}{white}{lightgray}
    \begin{tabular}{llrrrrrr}
        \toprule
         &  &  &  &  & \multicolumn{3}{c}{\textbf{Versus Baseline (Natural + FedAvg)}} \\
        \cmidrule(lr){6-8}
        \textbf{Scenario} & \textbf{Agg.} & \textbf{High x Low} & \textbf{High x Mid} & \textbf{Low x Mid} & $\Delta_{High,Low}$ & $\Delta_{High,Mid}$ & $\Delta_{Low,Mid}$ \\
        \midrule
        Natural & FedAvg  & 10.94 & 7.56 & 3.85 & -  & -  & -  \\
        Exclude Low & FedAvg  & 11.69 & 8.34 & 3.92 & -6.93\% & -10.30\% & -1.74\% \\
        Balanced & FedAvg  & 8.63  & 6.52 & 2.60 & \textbf{21.09\%} & 13.68\% & \textbf{32.49\%} \\
        Mildly tiered & FedAvg  & 13.56 & 8.69 & 5.30 & -23.95\% & -15.00\% & -37.74\% \\
        Fully tiered & FedAvg  & 13.12 & 8.13 & 5.35 & -19.96\% & -7.59\% & -39.01\% \\
        \hline
        Natural & FedProx & 9.87  & 6.30 & 3.88 & 9.72\%  & \textbf{16.60\%} & -0.87\% \\
        Exclude Low & FedProx & 10.58 & 7.65 & 3.46 & 3.21\%  & -1.22\% & 10.04\% \\
        Balanced & FedProx & 9.70  & 6.94 & 3.23 & 11.34\% & 8.23\%  & 16.14\% \\
        Mildly tiered & FedProx & 12.03 & 8.25 & 4.29 & -10.05\% & -9.11\% & -11.49\% \\
        Fully tiered & FedProx & 10.98 & 7.61 & 3.85 & -0.41\% & -0.74\% & 0.11\% \\
        \bottomrule
        \rowcolor{white}
        \multicolumn{4}{l}{\scriptsize Note: All comparisons have $p<0.05$ and represent statistical significance.} \\
    \end{tabular}
\end{table*}

We now present the results of our evaluation, key insights and discussion. \Cref{tab:agg_results,tab:fairness-dunn} report the fairness and accuracy results for all experiments relative to the baseline configuration (Natural distribution with FedAvg), which represents a deployment without awareness of device-induced data heterogeneity. We analyze how silo construction and aggregation strategies affect both predictive performance and inter-tier fairness in federated next-location prediction.

\subsection{Central model evaluation}

We first evaluate how a central model performs, testing the drift in performance when we later move for a federated system, serving as an upper-bound reference for accuracy and fairness. As expected, the centralized setting achieves the best accuracy ($273.75\,m$ median error), outperforming all federated configurations by 8--19\%. However, its fairness is $-6.87\%$ worse than the federated baseline. This result highlights that fairness degradation is primarily driven by bias in client and data distribution, rather than simply being federated.

\textbf{Takeaway.}
Centralization achieves best accuracy but worst fairness, confirming that fairness degradation is driven by client/data imbalance.

\subsection{Exclusion of low-tier devices}

We next investigate how excluding Low-tier devices from training can impact the overall fairness (keeping only High- and Mid-tiers). This is a common client-selection technique to avoid stragglers, which are common towards low-tier devices. However, as observed in previous works, this can lead to significant fairness implications for the excluded devices~\cite{maeng2022towards}.

Excluding Low-tier devices from training leads to a $-16.23\%$ reduction in fairness under FedAvg, while accuracy remains largely unchanged. Dunn test results show larger pairwise gaps across all tier comparisons, indicating uniformly larger performance gaps. This confirms that removing slower clients from training introduces substantial fairness risks for underrepresented devices.

\textbf{Takeaway.}
Excluding Low-tier devices to mitigate stragglers causes fairness collapse (-16.23\%), with statistically significant pairwise gaps. Systems optimizations against stragglers may directly conflict with the fairness goal.

\subsection{Balancing the distribution of tiers}
\label{subsec:balancing_tiers}

The natural distribution in our dataset is heavily biased towards High-tier devices, with the other two tiers underrepresented. The subsequent question is whether explicitly accounting for tiers, their imbalance and the different patterns they generate, can improve client selection across silos and lead to a fairer global model. 

Balancing the distribution of tiers across silos produces the largest fairness improvement ($+34.49\%$) with no meaningful degradation in global accuracy. We also observe the best gains in the disparity of pairwise tiers on the Dunn tests (21.09\% for $\Delta_{High,Low}$ and 32.49\% for $\Delta_{Low,Mid}$). This demonstrates that simple tier-aware client selection can substantially reduce inter-tier performance gaps. Notably, this improvement is achieved despite the remaining imbalance in trajectory length across tiers, suggesting that client-level balancing alone is sufficient to significantly improve fairness. It also goes against previous approaches that gave preference for selecting High Entropy clients for improving accuracy~\cite{park2023fedgeo}, showing that both fairness and accuracy can be gained by balancing selection across mobility profiles (in our case, through balancing smartphone tiers).

\textbf{Takeaway.}
Tier-aware client selection achieves +34.49\% fairness improvement without significant accuracy loss, showing that the fairness-accuracy trade-offs stem from data imbalance caused by system-level heterogeneity.

\subsection{Specialized silos}

We next explore if creating specialized silos impacts fairness, \ie whether each silo is composed of a single tier. This is motivated by prior work showing that training federated next-location prediction models benefited from having silos specialized in specific mobility tiers \cite{esper2024beauty}. Considering the similarities we observe between mobility clusters and smartphone tiers (\Cref{fig:sankey_clusters}), we aim to check if this holds true for our dataset, as well as how it impacts fairness. 

Tier-specialized silos degrade both accuracy and fairness. Fully tiered silos yield the largest accuracy drop ($-2.11\%$), while mildly tiered silos produce the largest fairness degradation ($-48.32\%$), which is counter-intuitively worse than the fully tiered case ($-36.99\%$). In the fully tiered setting, each silo learns a coherent local model specialized to one tier; while aggregation is challenging, gradient directions within each silo are internally consistent. In the mildly tiered setting, each silo contains a dominant tier mixed with minorities of the other two, generating conflicting gradient directions that neither full specialization nor full mixing can resolve. 
Homogeneous silo composition therefore appears preferable for both performance and fairness.

\textbf{Takeaway.}
Tier-specialized silos underperform in our settings, most likely due to conflicting gradient directions during model aggregation.

\subsection{Changes in model aggregation}
\label{subsec:model_agg}

A common solution to address in \ac{FL} data heterogeneity is to use a model aggregation technique that takes into account the existence of heterogeneity across clients and their resulting model updates. To address this solution, we re-train all models across experiments using FedProx as the model aggregation technique, evaluating its ability in solving the issue previously observed with FedAvg. 

Replacing FedAvg with FedProx mitigates fairness disparities across most experiments. Under the natural distribution, FedProx improves fairness by 21.46\% while also increasing accuracy by 1.05\%. The performance of using FedProx while also balancing the tiers is less effective (19.94\%) than its counterpart with FedAvg. This shows that while FedProx can be an alternative solution to mitigate fairness issues across smartphone tiers when this information is not available, it's not as effective as the solution from \Cref{subsec:balancing_tiers} (34.49\%)
This is highlighted by the pairwise Dunn tests from \Cref{tab:fairness-dunn}, where balancing the tiers together with FedAvg achieves more consistent results in the inter-tier fairness than the FedProx experiments, especially for $\Delta_{Low,Mid}$ where FedProx in a scenario with no knowledge of tiers (Natural) failed to reduce the fairness gap.

\textbf{Takeaway.}
FedProx (+21.46\% fairness without device info) is a practical fallback when tier metadata is unavailable, but tier-aware selection remains significantly better (+34.49\%) overall and present more consistensy in inter-tier fairness, quantifying the value of device-tier awareness.

\subsection{Discussion}


\subsubsection{Key findings and limitations.}
We study fairness under device-induced data heterogeneity in federated next-location prediction and identify two main findings. First, simple tier-aware client selection significantly improves fairness with only a negligible impact on accuracy. Second, when device-tier information is unavailable, heterogeneity-aware methods such as FedProx provide an effective alternative. Importantly, these fairness gains do not lead to meaningful accuracy degradation, challenging the commonly assumed fairness–accuracy trade-off in this setting and suggests that improvements mainly arise from addressing data imbalance rather than constraining optimization. Finally, although the smartphone ecosystem evolves rapidly due to frequent device releases, our analysis shows stable tier distributions between 2022 and 2024; nevertheless, we recommend periodically updating the tiering and clustering to ensure the continued validity of these assumptions.

\subsubsection{Impact on future deployment scenarios.}
Beyond standard next-location prediction, our device-tier-aware findings transfer to three emerging mobile network scenarios:

\begin{list}{\textbullet}{
  \setlength{\leftmargin}{0pt}
  \setlength{\itemsep}{0pt}
  \setlength{\topsep}{0pt}
  \setlength{\itemindent}{1em}
  }
    \item In \textit{roaming} contexts~\cite{fezeu2024roaming}, visiting devices may introduce additional tier heterogeneity as international visitors often carry different device distributions than the home network population. FL systems should account for this dynamic tier composition during client selection to maintain fairness across both local and roaming users.
    
    \item For \textit{predictive network orchestration}~\cite{farooq2020mobility,ma2020survey,kalntis2025smooth}, device tiers can inform proactive resource allocation: High-tier devices with higher mobility patterns (\Cref{sec:mobility_analysis}) may benefit from predictive dual connectivity or advanced handover optimization, while Low-tier devices with constrained mobility require different network management strategies.
    
    \item In \textit{AI edge offloading}~\cite{wu2019mobile,zhang2023trajectory}, device tiers can guide model partitioning decisions, with High-tier devices running larger on-device models and Low-tier devices offloading computation to edge servers. FL fairness strategies must adapt to ensure equitable performance across this hybrid deployment model, where computation capabilities directly influence both training participation and inference quality.
\end{list}

\section{Conclusions}
\label{sec:conclusion}



This work conducted a large-scale measurement study of device-induced mobility heterogeneity using handover data from a major \ac{MNO}. Across over $50{,}000$ \acp{UE} and 178 smartphone models, we show that device tiers are strongly associated with mobility, revealing a systematic source of data imbalance. 
We then demonstrate the downstream impact on \ac{FL} next-location prediction: unbalanced tier distributions degrade inter-tier fairness, even when aggregate accuracy appears competitive. Tier-aware client balancing is the most effective strategy for improving fairness (+34.49\%) with no meaningful loss in accuracy. When tier metadata is unavailable, heterogeneity-aware aggregation (FedProx) offers a practical alternative (+21.46\%).


Our study has a set of limitations. We focus on a single city in a wealthy European country during a 14-day period and our insights may not generalize directly to national deployments or in different countries. Additionally, users upgrade phones and tier distributions shift over time, requiring a periodic re-evaluation to ensure the relevance of our findings.

In future works, we plan to address these limitations to validate tier stability and understand device adoption cycles, as well as expand the geographic analysis to assess whether the device–mobility–fairness across space.

\bibliographystyle{ieeetr}
\bibliography{refs}

\appendix
\section{Supplementary material}

\subsection{Time encoding}
\label{subsec:time_encoding}

Because time of day is periodic, representing it as a scalar would introduce artificial discontinuities (\eg 23:00 and 00:00 appear far apart). To preserve temporal continuity, we encode~\cite{feng2018deepmove} the time since midnight as a fraction of the day and project it onto the unit circle using sine and cosine transformations, yielding a two-dimensional cyclical feature. 

We do this as follows: we first represent the timestamp of a trajectory as a fraction between the number of seconds of that timestamp from midnight over the total number of seconds in a day ($86,400$s). For example, the timestamp 14:30 would be $52,200$s, resulting in a fraction of the day of $t_{frac}=0.6042$. With this, we will calculate the both the sine and cosine components of this fraction as: 

\begin{align}
    h_{\text{sin}} &= \sin(2\pi t_{\text{frac}}) \\
    h_{\text{cos}} &= \cos(2\pi t_{\text{frac}})
\end{align}

\noindent with $2\pi t_{\text{frac}}$ representing the angle in radians. So, for the 14:30 timestamp, we would obtain as input features of the model $(h_{\text{sin}},\, h_{\text{cos}}) \approx (-0.61,\,-0.79)$. 

\subsection{Haversine distance}
\label{subsec:haversine}

To capture user mobility dynamics, we compute the distance traveled between consecutive handovers. Given two successive geographic coordinates $[(\phi_{1},\lambda_{1}),(\phi_{2},\lambda_{2})]$, we estimate the great-circle distance using the Haversine formula, which accounts for the Earth’s curvature and is commonly used in mobility prediction~\cite{feng2020pmf,zhang2023federated}. Distances are expressed in kilometers in the formula, and the first point of each trajectory is assigned a distance of zero. We compute the distance as:

\begin{align}
    a &= \sin^2\!\left(\frac{\Delta \phi}{2}\right) 
        + \cos(\phi_1)\cos(\phi_2)\sin^2\!\left(\frac{\Delta \lambda}{2}\right) \\
    d &= 2R \arctan2(\sqrt{a}, \sqrt{1-a})
\end{align}

\noindent where $\Delta \phi = \phi_2 - \phi_1$, $\Delta \lambda = \lambda_2 - \lambda_1$,
and $R$ is the Earth's radius (6371 km).











\section{Supplementary results}




\subsection{Silhouette score analysis}
\label{sec:silhouette}

To select the number of clusters used for device-tier grouping, and later for the mobility clusters, we evaluate the clustering quality across multiple values of $k$ using the Silhouette score~\cite{rousseeuw1987silhouettes}. This metric measures how well each sample fits within its assigned cluster compared to neighboring clusters, and ranges from $-1$ to $1$, where higher values indicate better-separated and more cohesive clusters.

\begin{figure}
    \centering
    
    \begin{subfigure}[t]{0.32\linewidth}
    \centering
        \includegraphics[width=\linewidth]{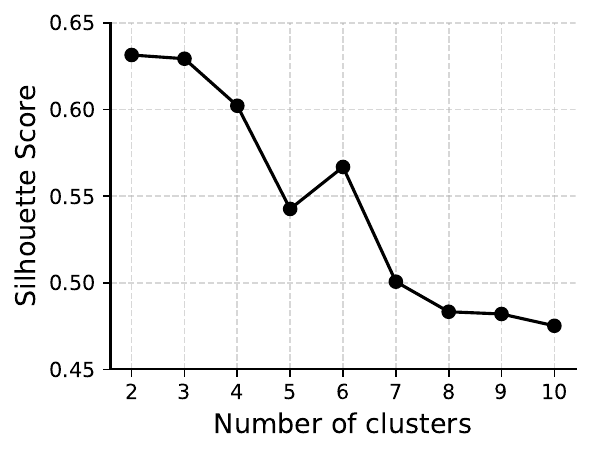}
        \caption{Devices}
        \label{fig:silhouette_smartphones}
    \end{subfigure}
    \hfill
    \begin{subfigure}[t]{0.32\linewidth}
        \centering
        \includegraphics[width=\linewidth]{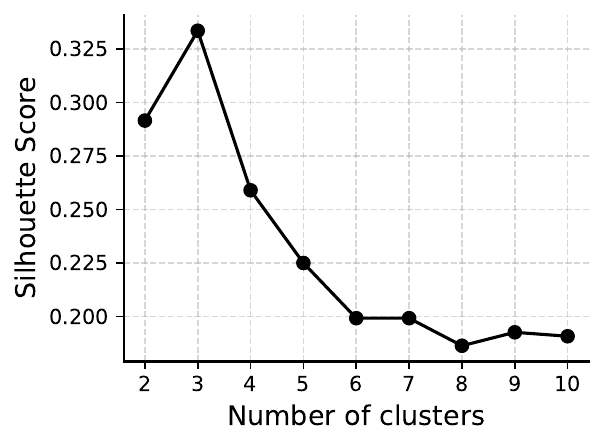}
        \caption{Entropy}
        \label{fig:silhouette_entropy}
    \end{subfigure}
    \hfill
    \begin{subfigure}[t]{0.32\linewidth}
        \centering
        \includegraphics[width=\linewidth]{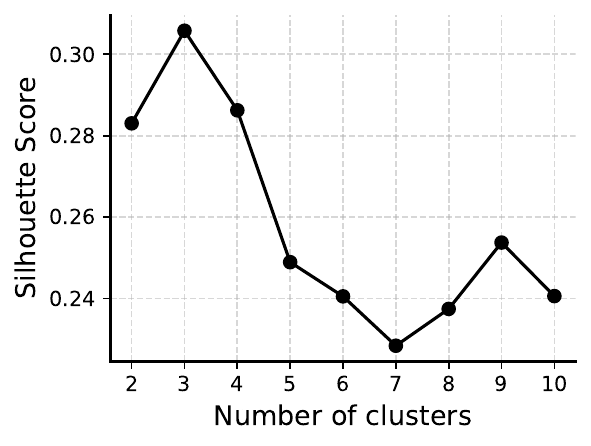}
        \caption{RoG}
        \label{fig:silhouette_rog}
    \end{subfigure}
    \caption{
    Silhouette scores obtained from the k-means clustering for the (a) Devices, (b) Entropy and (c) RoG tiers.
    }
    \label{fig:silhouette_analysis}
\end{figure}

We compute the average Silhouette score for values of $k$ in the range $[2,10]$ for the three clustering tasks used to derive the Device, Entropy, and RoG tiers. As shown in \Cref{fig:silhouette_analysis}, the configuration with $k=3$ achieves the highest score before a marked decline in clustering quality. This indicates that increasing the number of clusters beyond three reduces cluster separability and quality, leading to less well-defined groupings.

\subsection{Nationwide device model population (2022)}

\begin{figure}
    \centering
    
    \begin{subfigure}[t]{\columnwidth}
        \centering
        \includegraphics[width=\linewidth]{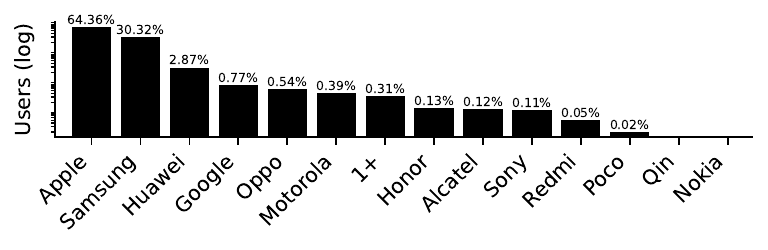}
        \caption{}
        \label{fig:brand_distribution_2022}
    \end{subfigure}
    
    \vspace{0.6em}
    
    \begin{subfigure}[t]{\columnwidth}
        \centering
        \includegraphics[width=\linewidth]{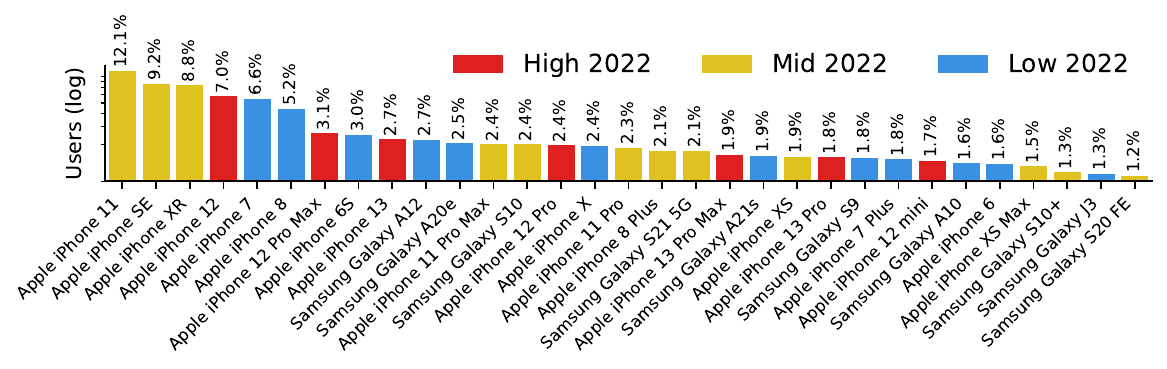}
        \caption{}
        \label{fig:device_distribution_2022}
    \end{subfigure}
    
    \caption{(a) Brand and (b) device distribution for the entire country. Tiers represented here involve only devices released by 2022 (the year this set was collected).}
    \label{fig:plots_country_2022}
\end{figure}

We utilize a dataset that catalogs all devices connected to the network of the operator, including their TAC values, to understand how the entire country's device population is composed, and how it compares to the capital's sample. 
Unfortunately this catalog had its last version collected in 2022, so we re-did the tier-clustering from \Cref{sec:system_analysis} for \Cref{fig:plots_country_2022} to account for only 2022 devices (as we use age as a feature and results from the 2024 population would be skewed). 

As it can be seen on \Cref{fig:brand_distribution_2022}, the share across brands is similar to 2024. However, as previously mentioned, the smartphone tiers change, skewing towards Mid- and Low-tier devices for the entire catalog (even with the 2022 clusters). This can be explained by the country's catalog having still older devices from the dominant brands: while in 2022 the latest released iPhone was the 13 series, the country still had a significant population of Series 11 and below models, and less of a presence of Pro and Pro Max versions (which were more present in the capital's 2024 set). 


\subsection{Index of Deprivation Rank distribution across device tiers and mobility clusters}

\begin{figure}
    \centering
    
    \begin{subfigure}[t]{0.32\linewidth}
        \centering
        \includegraphics[width=\linewidth]{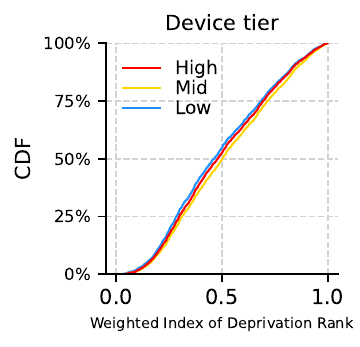}
        \caption{}
        \label{fig:income_tier_cluster_cdf}
    \end{subfigure}
    \hfill
    \begin{subfigure}[t]{0.32\linewidth}
        \centering
        \includegraphics[width=\linewidth]{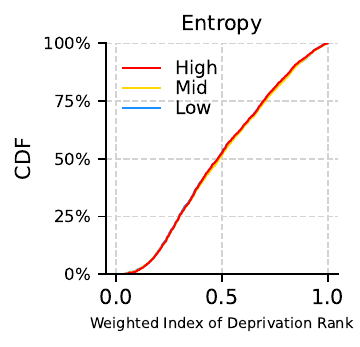}
        \caption{}
        \label{fig:income_entropy_cluster_cdf}
    \end{subfigure}
    \hfill
    \begin{subfigure}[t]{0.32\linewidth}
        \centering
        \includegraphics[width=\linewidth]{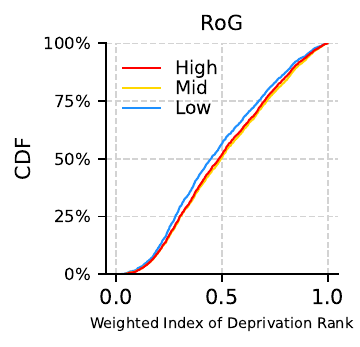}
        \caption{}
        \label{fig:income_rog_cluster_cdf}
    \end{subfigure}
    
    \caption{Distribution of the Index of Deprivation Rank for \ac{UE} belonging to each High/Mid/Low class of (a) Device tiers, (b) Entropy and (c) RoG clusters. Smaller values of the Index of Deprivation Rank indicate that the area that an UE is located is more deprivation than others.}
    \label{fig:deprivation_cdfs}
\end{figure}

Considering the place of residence of \acp{UE} belonging to each device model, and considering the device tier, Entropy and \ac{RoG} class of each device, we obtain the distribution of the Index of Deprivation Rank for each of those populations (with lower values indicating higher deprivation), with the goal of understanding if the deprivation experienced at the home location of \acp{UE} have a correlation with the categories discussed throughout this work. Besides the \ac{KS} tests presented in \Cref{tab:income_ks_test}, we also present the CDFs of the Index of Deprivation Rank across these classes in \Cref{fig:deprivation_cdfs}. We particularly observe that \acp{UE} from the Low device and Low RoG tiers tend to have a lower rank, indicating that their home location suffers from higher deprivation than what's experienced by \acp{UE} at the respective High and Mid tiers. 


\subsection{Model and loss function evaluation}
\label{subsec:model_loss_eval}

We report on \Cref{tab:model_performance} the performance of the different model architectures and loss functions evaluated for centralized next-location prediction. The target here was to find the best performing model for the centralized setting, before moving to the federated setting. We observe that moving from NMSE to Haversine distance as loss function improves the accuracy of all models, and that the RNN~\cite{liu2016predicting} architecture trained with the Haversine loss achieves the best performance across all over the LSTM~\cite{kong2018hst} and Transformer-based alternatives. These results drive our decision to move forward with only the RNN architecture for the \ac{FL} evaluation.

\begin{table}[t]
    \caption{Model performance comparison across architectures and loss functions.}
    \centering
    \rowcolors{2}{white}{lightgray}
    \begin{tabular}{lllll}
        \toprule
        \textbf{Model} & \textbf{Loss} & \textbf{$\mu$ (m)} & \textbf{Med. (m)} & \textbf{ACC@1} \\
        \hline
        LSTM        & NMSE       & 753.73 & 434.47 & 26.43\% \\
        LSTM        & Haversine  & 740.71 & 310.57 & 35.89\% \\
        Trans. & NMSE       & 840.57 & 533.62 & 20.08\% \\
        Trans. & Haversine  & 750.73 & 350.30 & 33.01\% \\
        RNN         & NMSE       & 826.02 & 510.86 & 21.21\% \\
        RNN         & Haversine  & \textbf{737.24} & \textbf{273.75} & \textbf{39.06\%} \\
        \bottomrule
    \end{tabular}
    \label{tab:model_performance}
\end{table}



\end{document}